\documentclass[11pt,letterpaper,preprint,floatfix,amsmath,onecolumn]{revtex4}

\usepackage{amsmath}
\usepackage{epsfig}
\usepackage{array}
\usepackage{verbatim}
\usepackage{hyperref}
\usepackage{amssymb}

\newcommand{\re}[1]{(\ref{#1})}

\newcommand{\beg}{\begin{equation}}

\newcommand{\en}{\end{equation}}

\newcommand {\dis}{\displaystyle}

\newcommand{\eps}{\varepsilon}
\newcommand{\lam}{\lambda}

\newcommand{\w}{\omega}

\newcommand{\eref}[1]{Eq.~(\ref{#1})}

\begin{document}

\title{The Link between Integrability,  Level Crossings, and Exact Solution in Quantum  Models}

\author{H. K. Owusu, K. Wagh, and E. A. Yuzbashyan}

\affiliation{Center for Materials Theory, Department of Physics and Astronomy,
Rutgers University, Piscataway, NJ 08854, USA}

\begin{abstract}
We investigate the connection between energy level crossings in
integrable systems and their integrability, i.e. the existence of
a set of non-trivial integrals of motion. In particular, we
consider a general quantum Hamiltonian linear in the coupling $u$,
$H(u)=T+uV$, and require that it    have  the maximum possible
number of nontrivial  commuting partners also linear in $u$. We
demonstrate how this commutation requirement {\it alone} leads to:
1) an exact solution for the energy spectrum and 2) level
crossings, which are {\it always} present in these Hamiltonians in
 violation of the  Wigner-von Neumann
non-crossing rule. Moreover, we
construct these Hamiltonians explicitly by resolving the above
commutation requirement and show their equivalence to a sector of
Gaudin magnets (central spin Hamiltonians). By contrast,
fewer than the maximum number of conservation laws does not
guarantee level crossings.
\end{abstract}
\date{\today}
\maketitle

%\pagenumbering{Roman}

\tableofcontents

\newpage

%\pagenumbering{arabic}

\section{Introduction}

Level crossings -- the emergence of degeneracies in a physical
system at a certain value of some tuned system coupling -- underly
a myriad of compelling phenomena,   including anomalies in
 relaxation rates\cite{Happer},
   the onset of quantum chaos\cite{chaos}, quantum phase transitions\cite{Sachdev},
   Berry's phase\cite{Berry,Bhattacharya} etc. It is widely believed that these
degeneracies can often be understood in terms of a certain
underlying symmetry. However, in many cases this connection
between symmetry and degeneracy remains mysterious. This is
especially true for quantum  integrable systems,  e.g. the 1d
Hubbard, anisotropic Heisenberg,  reduced BCS models etc. These
systems are long known to display an abundance of level
crossings[\citealp{hl}--\citealp{strong}], see Fig.~\ref{fig: typical
integrable}, in violation of the famous  Wigner-von Neumann
non-crossing rule[\citealp{Hund}--\citealp{Kestner and Duan}] and with no convincing symmetry
explanation.

In this paper we {\it derive}   the existence of level crossings and   an exact solution
for a general parameter-dependent quantum Hamiltonian {\it from}
its integrability.  Our work has been inspired in part by Refs.~\onlinecite{Gaudin} and \onlinecite{emil} and especially Shastry's paper \cite{Shastry}, which opened up a new, purely algebraic  perspective on quantum integrable models independent of Bethe's Ansatz. In Hamiltonian mechanics the integrability of
a system with $n$ degrees of freedom is usually understood as the
existence of a maximum number ($n$) of Poisson commuting
independent invariants. Then, a well-known theorem due to
Liouville and Arnold guarantees that  the equations of motion can be solved
 by quadratures\cite{arnold}. There is no similarly accepted notion
of quantum integrability, especially in finite dimensional
systems, e.g. discrete lattice models in condensed matter physics
where the state space is generally finite. In particular, it is
often unclear what constitutes an independent integral and what is
the natural notion of the number of degrees of freedom.
Nevertheless, it turns out that these difficulties can be
circumvented if one restricts the  manner in which the integrals of
motion depend on the  coupling.

For concreteness, let us consider Hamiltonians
linear in the coupling $u$. As we are interested in discrete
energy spectra, we assume that the Hamiltonian can be represented
by an $N\times N$ matrix.   Following the classical notion of
integrability, we require  the existence of the {\it maximum} possible number of
independent (see below) mutually commuting integrals, $[H^i(u),
H^j(u)]=0$,
 where  $H^i(u)=T^i+uV^i$ are Hermitian operators. One of them is
 the Hamiltonian itself, e.g. $H^1(u)\equiv H(u)$. Using this
commutation requirement {\it alone}, we  derive an exact solution
for the spectrum of each $H^i(u)$, which can be viewed as an
extension of the  Liouville-Arnold theorem to quantum Hamiltonians.
Moreover, we are able to demonstrate that the eigenvalues  of
$H^i(u)$ are necessarily degenerate at a discrete set of values of
$u$.

First, we solve the nonlinear commutation relations $[H^i(u),
H^j(u)]=0$ to obtain each $H^i(u)$ explicitly, see below.
Interestingly, it turns out that   these  {\it maximally commuting}
(or simply maximal) operators $H^i(u)$  can be mapped to exactly
solvable Gaudin magnets \cite{Gaudin,Ushveridze} (central spin
Hamiltonians). The latter describe a localized spin in a magnetic
field $B=u$ interacting with $N-1$ ``environmental'' spins and
have a variety of physical
applications~[\citealp{subir}--\citealp{lukin}]. The mapping to
Gaudin magnets allows us to obtain the exact solution for the
eigenvalues and eigenfunctions of $H^i(u)$ and analyze its level
crossings. The total number of pairwise crossings  varies as
$M_c=(N-1)(N-2)/2-2K>0$, where $K$ is a positive integer such that $M_c\ge1$.  For
instance, $N=5$  maximal operators display $2$, 4, and 6 instances
of level crossings. On the other hand, we demonstrate that
Hamiltonians having less than the maximum number of $u$-dependent
integrals can have no level crossings.   Further, we show that the
coupling-dependent commuting matrices obtained within the approach
developed by Shastry in Ref.~\onlinecite{Shastry} (where it was
also conjectured that these matrices always have crossings, see
also Ref.~\onlinecite{emil})  are maximal, even though our
constructions are quite different.

Pairwise crossings of energy levels  are
usually understood in the context of the Wigner-von Neumann
non-crossing rule. This rule initially suggested by
Hund\cite{Hund} and justified by Wigner and von Neumann\cite{von
Neumann}  has thereafter seen restatements and refinements by a
number of authors[\citealp{Teller}--\citealp{Kestner and
Duan}]. It states that eigenstates of the same symmetry do not
cross as a function of a single coupling parameter. This can be
seen, for example, from the following argument. Suppose two energy
levels $E_{1}\left(u\right)$ and $E_{2}\left(u\right)$ of
$H\left(u\right)$ come close at a certain $u=u_{0}$. Expanding in
a vicinity of $u_{0}$: $H\left(u\right)\approx
H\left(u_{0}\right)+\left(u-u_{0}\right)V\left(u_{0}\right)$ and
using ordinary perturbation theory, we obtain \cite{Pechukas}
\begin{equation}
\frac{d^{2}\Delta}{du_{0}^{2}}=\frac{4V_{12}^{2}\left(u_{0}\right)}{\Delta\left(u_{0}\right)}+
F\left(u_{0}\right)
\label{eq: pert. repulsion}
\end{equation}
 where $\Delta\left(u_{0}\right)=E_{1}\left(u_{0}\right)-E_{2}\left(u_{0}\right)$,
$V_{12}\left(u_{0}\right)$ is the matrix element of the
perturbation $V\left(u_{0}\right)$ between states
$\left|1\right\rangle $ and $\left|2\right\rangle $, and
$F\left(u_{0}\right)$ represents the contribution of the remaining
states. We see from Eq.~\eqref{eq: pert. repulsion} that as the
two levels approach, $\Delta\left(u_{0}\right)\rightarrow0$,
infinite repulsion sets in, preventing them from crossing. This is
indeed what takes place in the absence of symmetry -- energy levels
repel see Fig.~\ref{fig: avoided crossings}. The situation changes if the Hamiltonian $H\left(u\right)$ possesses
a $u$-independent symmetry $S$, i.e. $\left[H\left(u\right),S\right]=0$.
This can be a spatial rotation, translational invariance,
internal space reconfiguration, etc. Because $S$ does not depend
on the coupling $u$, it  commutes with $H\left(u_{0}\right)$
and $V\left(u_{0}\right)$ individually. Evaluating the matrix element of $\left[V\left(u_{0}\right),S\right]$
between states $\left|1\right\rangle $ and $\left|2\right\rangle $,
we obtain $V_{12}\left(u_{0}\right)=0$ for any $u_{0}$ as long as
 $\left|1\right\rangle $ and $\left|2\right\rangle $ have
different symmetry,  $s_{1}\neq s_{2}$, where $S\left|1\right\rangle =s_{1}\left|1\right\rangle $
and $S\left|2\right\rangle =s_{2}\left|2\right\rangle $. Thus, while
levels of different symmetry can cross, crossings of levels of the
same symmetry are prohibited.

\begin{figure}
\begin{centering}
\includegraphics[scale=0.8]{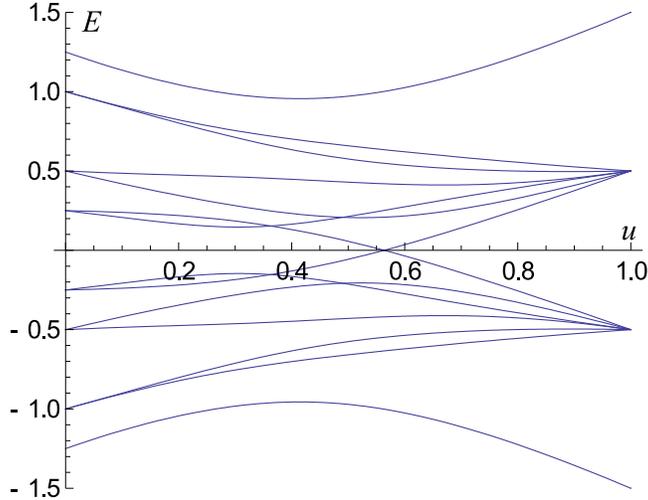}
\end{centering}
\caption{\label{fig: typical integrable} Energies of 1d Hubbard model on six cites characterized
by a complete set of quantum numbers, i.e. all levels have the same
$u$-independent symmetry, see Refs.~\onlinecite{hl} and \onlinecite{emil}. The energies are in units of $U-4T$ plotted
as functions of $u=U/(U-4T)$, where $U$ is
the strength of the Coulomb repulsion and $T<0$ is the hopping matrix
element. The parameter $u$ varies from 0 to 1 as $U$ goes
from $0$ to $\infty$. Note that, in violation of the Wigner-von
Neumann non-crossing rule, we see a profusion of level crossings for
states of the same symmetry.}
\end{figure}

\begin{figure}
\begin{centering}
\includegraphics[scale=0.8]{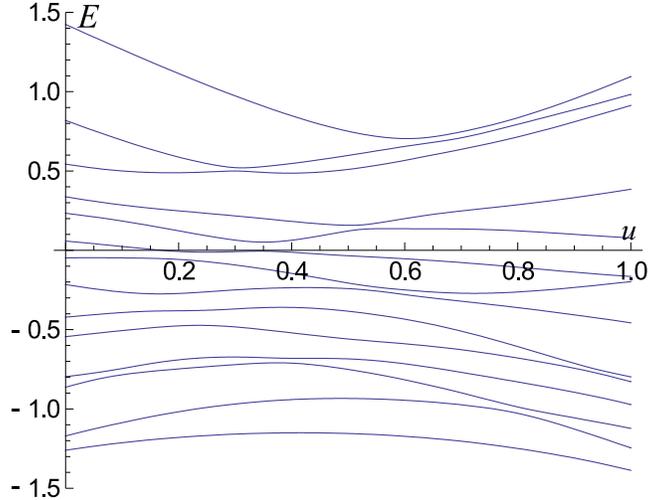}
\end{centering}
\caption{\label{fig: avoided crossings} Energy levels of a $14\times 14$
 Hamiltonian $H\left(u\right)=A+u\, B$, where independent matrix elements
of $A$ and $B$ are  uniformly distributed random  numbers. Note
that though levels
do approach one another closely, they never cross. A typical Hamiltonian
with no $u$-independent symmetry exhibits such level repulsion, see \eref{eq: pert. repulsion},
in contradistinction to what takes place in integrable systems, see Fig. \ref{fig: typical integrable}. Here
and throughout this paper we use  {\it Mathematica} program to perform numerical tests and plot the results.}
\end{figure}

Unfortunately, this basic argument does not extend to quantum
integrable Hamiltonians $H\left(u\right)$, which   typically
violate the non-crossing rule. Indeed, these systems show   crossings
of energy levels that have the same quantum numbers for
all $u$-independent symmetry[\citealp{hl}--\citealp{strong}], see e.g. Fig.~\ref{fig:
typical integrable}. Integrable Hamiltonians are known to have
special coupling dependent conserved currents, ``dynamical
symmetries'', in addition to $u$-independent symmetries. It is
tempting to attribute these crossings to such symmetries. On the
other hand, it is crucial for the validity of the non-crossing
rule that the symmetry $S$ be $u$-independent. Indeed, consider an
integrable Hamiltonian acting on a finite-dimensional  space, e.g.
a lattice model with a finite number of sites. Let
$H\left(u\right)$ be one of its blocks characterized by the same
quantum numbers for a complete set of mutually commuting
$u$-independent symmetries and let $\widetilde{H}\left(u\right)$
be the corresponding block of one of the conserved currents
\begin{equation}
[H(u),\widetilde{H}(u)]=0\quad\mbox{for all $u$}.
 \label{comcur}
\end{equation}
 Due to the $u$-dependence, $\widetilde{H}\left(u_0\right)$ no longer
commutes with $V\left(u_0\right)$ separately and, therefore, the
above argument lifting the level repulsion does not hold. At the
same time, given a crossing one can always artificially engineer a
 ``conserved current'' that commutes with
$H\left(u\right)$. Therefore, restrictions on the form of
$\widetilde{H}\left(u\right)$ are necessary to make meaningful
contact between symmetries and degeneracies.

To be specific, let $H(u)=T+uV$ and
$\widetilde{H}(u)=\widetilde{T}+u\widetilde{V}$ be Hermitian
operators acting on an $N$-dimensional  space, i.e. they can be
represented by $N\times N$ matrices. Eq.~\eqref{comcur} implies
\begin{equation}
[T,\widetilde{T}]=[V,\widetilde{V}]=0,\qquad[T,\widetilde{V}]=[\widetilde{T},V].
\label{comrel}
\end{equation}
 For any linear $H\left(u\right)$ there are always trivially related
commuting partners $\widetilde{H}\left(u\right)=aH\left(u\right)+\left(b+c\, u\right)I$, where $I$ is an identity matrix. However,
the requirement that \eref{comrel}   have nontrivial solutions  leads to
a set of nonlinear constraints that severely restrict the matrix elements
of both $H(u)$ and $\widetilde{H}(u)$. For example, for $N=3$ eliminating  $\widetilde{T}$ and $\widetilde{V}$
from \eref{comrel}, one obtains a single nonlinear constraint on the matrix elements of $H(u)$ \cite{emil}. In view
of the   preceding discussion   regarding the prevalence of level crossings in integrable models,  a natural question
is whether these constraints, i.e. the existence of a nontrivial    $\widetilde{H}(u)$, imply crossings in the spectrum
of $H(u)$ and vice versa. This is indeed the case for $N=3$. Specifically, one can show that $3\times3$ matrices $H(u)=T+uV$
that have nontrivial commuting partners also have a  level crossing and vice versa\cite{emil}. However,  this is no longer true
for $N\ge 4$ -- \eref{comrel}
does not necessarily lead to level crossings.   Moreover, crossings occur even in the absence of nontrivial
 partners and $u$-independent symmetries, see below.
We see that  a single dynamical symmetry is insufficient to
explain level crossings. On the other hand, quantum integrable
Hamiltonians typically have more than one coupling dependent
commuting operator.  In fact, as we show below, the maximum
possible number -- which turns out to be $N$ -- of integrals are
necessary to ensure level crossings.

We define the set of maximally commuting Hamiltonians as a vector
space, $\mathcal{M}$, formed by $N\ge3$ Hermitian, mutually
commuting $N\times N$ matrices $H^{i}\left(u\right)=T^{i}+u\,
V^{i}$ together with the $N\times N$ identity matrix $I$,
\begin{equation}
\left[H^{i}\left(u\right),H^{j}\left(u\right)\right]=0,\mbox{ for
all $u$ and $i,j=1,\dots,N,$}
 \label{alcom}
\end{equation}
where $u$ is a real parameter. Operators $H^{i}(u)$ are assumed to
be independent in that matrices $V^i$ are linearly independent,
i.e. $\sum_{i=1}^N c_i V^i=0$  iff $c_i=0$ for all $i$
(equivalently one can require that $T^i$ be linearly independent).
In addition,  $H^{i}\left(u\right)$ are taken to have no
$u$-independent symmetry common to all $H^{i}\left(u\right)$\cite{note2},
\beg
\nexists\:\Omega\neq
a\,I \mbox{ such that }\;\left[\Omega,H^{i}(u)\right]=0 \mbox{ for all $u$ and $i$}.
\label{nou}
\en Therefore,
an arbitrary element $H(u)=T+uV$ of the vector space $\mathcal{M}$
has the form
\begin{equation}
H\left(u\right)=\sum_{i=1}^{N}d_{i} H^{i}\left(u\right) +a I,
\label{arbH}
\end{equation}
where $d_i$ and $a$ are real numbers. The addition of multiples of
the identity   affects neither commutation relations nor
level crossings of $H(u)$ and we will often omit the term $aI$ in
\eref{arbH}. Note also that \eref{arbH} implies that operators $H^i(u)$ together with $I$ provide a basis in the vector space
$\mathcal{M}$ of maximal Hamiltonians.

The set $\mathcal{M}$ is maximal in the sense that any Hermitian $H\left(u\right)=T+u V$ that commutes with all $H^i(u)$
can be written in the form \re{arbH}. Indeed, since $V$ and all
$V^i$ mutually commute, see \eref{comrel}, we can go to their
common eigenbasis. In this basis, the  $N$   diagonal matrices $V^i$ are
$N$ linearly independent $N$-dimensional vectors and, therefore,
there exist real numbers $d_i$ such that $V=\sum_{i=1}^N d_i V^i$.
The matrix $H(u)-\sum_{i=1}^N d_i H^i(u)$ is
$u$-independent and, since it also commutes with all $H^i(u)$, it must be
of the form $aI$ according to \eref{nou}.  Thus, $H(u)$ is of the form
\re{arbH}. By a similar argument one can show that one of the basic
matrices $H^i(u)$ can be chosen as $H^i(u)=(a+ub)I$ with real coefficients $a$ and $b$. We see that
there are $N-1$ nontrivial independent commuting operators. Therefore, the first nontrivial dimensionality is
$N=3$.

In what follows we begin with the explicit construction of a general, maximally commuting Hamiltonian $H(u)$. This is
done in Sec.~\ref{sec:Type-I-Basis} by choosing a convenient basis in the vector space $\mathcal{M}$ and solving \eref{alcom}.
In Sec.~\ref{algebra} we establish some useful algebraic properties of $H(u)$. Interestingly, it turns out that
the product of any two maximally commuting Hamiltonians can be written as a linear superposition of such Hamiltonians, i.e.
the set $\mathcal{M}$ has a certain quasi-ring structure.

Our parametrization of the maximally commuting Hamiltonians makes it transparent that they are related to
the Gaudin magnets \cite{Gaudin,Ushveridze},
see Sec.~\ref{sec: Central Spin Equiv.}. The latter are $N$ quantum spin Hamiltonians
\begin{equation}
\hat h^{i}\left(B\right)=-B\, \hat
s_i^z+\sum_{k=1}^{N}\lefteqn{\phantom{\sum}}'
\frac{\hat{\vec{s}}_{i}\cdot\hat{\vec{s}}_k}{\varepsilon_{i}-
\varepsilon_{k}}, \quad i=1,\dots,N, \label{gaud}
\end{equation}
 where the prime indicates that the summation is over $k\ne i$, $B$ is the $z$-aligned magnetic field, $\hat{\vec{s}}_{i}$ is an
 operator of spin
of magnitude $s_i$, and $1/(\varepsilon_{i}-\varepsilon_{k})$ is the coupling between spins $\hat{\vec{s}}_{i}$
and  $\hat{\vec{s}}_{k}$. The Hamiltonians $\hat h^i$ form a mutually commuting family
\beg
\left[\hat h^i(B), \hat h^j(B)\right]=0\quad\mbox{for any $B$}.
\label{gaudcom}
\en
Note also that each $\hat h^i(B)$ is invariant under rotations around the $z$ axis, which means
 the $z$ component of the total spin $\hat J^z=\sum_{i=1}^N \hat s_i^z$ is conserved
\beg
\left[\hat h^{i}, \hat J^z\right]=0, \quad i=1,\dots,N.
\label{Jz}
\en
As we will see in Sec.~\ref{sec: Central Spin Equiv.}, the maximally commuting Hamiltonians \re{arbH} correspond to
the sector of Gaudin magnets with $J^z=J^z_{\max}-1$, where $J^z_{\max}=\sum_{i=1}^N s_i$ is the maximum eigenvalue of $\hat J^z$.

In Sec.~\ref{sec: Exact Solution} we employ the mapping to the Gaudin magnets to obtain the exact solution for the
spectra of maximally commuting Hamiltonians. Using this solution, we analyze the asymptotic behavior of the eigenstates
in the limits $u\to\pm\infty$ in Sec.~\ref{sec: Crossings}. Matching the two limits allows us to establish the presence
of level crossings and to count them.   Next, we consider Hamiltonians having less than the maximum
number of commuting partners. In Sec.~\ref{submax},
we construct a family of $4\times 4$ Hermitian operators linear in the coupling $u$
that have a single nontrivial  partner as opposed to two partners for the $N=4$ maximal set.
It turns out that these submaximal Hamiltonians often display no degeneracies at all.
Finally, in Appendix we review Shastry's approach to constructing commuting real symmetric operators
and show that the resulting  operators are always maximal.

\section{\label{sec:Type-I-Basis}The Parametrization of maximally commuting Hamiltonians}

We begin our analysis by choosing a convenient basis in the vector
space  of maximally commuting Hamiltonians, which allows us to
 solve \eref{alcom} explicitly. The solution yields a
convenient parametrization for a general maximal
Hamiltonian \re{arbH} and elucidates the algebraic structure of
these operators. It also makes transparent the relationship
between these operators and Gaudin magnets~\re{gaud}.

Consider the maximal operators $H^i(u)=T^i+u V^i$
defined in Eqs.~\re{alcom} and \re{nou}. It follows from
\eref{alcom} that all $V^i$ commute with each other, see
Eqs.~\re{comrel} and \re{comcur}. By a $u$-independent unitary
 transformation we  go to a basis where all $V^i$ are diagonal. Since
$V^i$ are also linearly independent, we can bring them to the
following ``canonical'' form by a linear transformation
\begin{equation}
D^{k}=\sum_{i}v_{i}^{k}V^{i} \label{dv},
\end{equation}
where $v_{i}^{k}$ are real numbers, $D^k$ are diagonal
with a single nonzero  matrix element $[D^k]_{jj}\equiv
D^k_j=\delta_{jk}$.  Next, we introduce a
``canonical'' basis in the space of maximally commuting operators
\beg
h^{i}(u)=E^i+uD^i=\sum_{j=1}^{N}v_{j}^{i}H^{j}(u),\quad
i=1,\dots,N. \label{canbas}
\en
The operators $h^i(u)$ have all
the properties of maximally commuting Hamiltonians defined in
Eqs.~\re{alcom} and \re{nou} as long as $H^i(u)$ do. In
particular,
\begin{equation}
\left[h^{i}\left(u\right),h^{j}\left(u\right)\right]=0,\quad\mbox{for all $u$ and $i,j=1,\dots,N$}.
\label{smalh}
\end{equation}
It follows from \eref{arbH} that a general maximally commuting operator can be written as
\begin{equation}
H\left(u\right)\equiv T+uV=\sum_{k=1}^{N}d_{k} h^{k}\left(u\right) +a I,
\label{arbh}
\end{equation}
where $d_k$ and $a$ are  real numbers. Note that with our choice of $D^k$,
$d_k$ are the eigenvalues of $V$.

To determine $H(u)$ explicitly, we need to solve \eref{smalh}. In
terms of $D^i$ and $E^i$ these equations read
 \begin{equation}
\left[D^{i},D^{j}\right]=0,
\quad\left[D^{i},E^{j}\right]=\left[D^{j},E^{i}\right],\quad
\left[E^{i},E^{j}\right]=0.
 \label{eq: Analogue}
\end{equation}
The first equation holds since $D^i$ are diagonal. The second
equation in terms of matrix elements is
\begin{equation}
\left(D_{m}^{i}-D_{n}^{i}\right)E_{mn}^{j}=\left(D_{m}^{j}-D_{n}^{j}\right)E_{mn}^{i}
\label{eq: Vanishing E elements}
\end{equation}
 where   $E_{mn}^{j}$  is the   $mn^{\mathrm{th}}$ matrix element of  $E^{j}$. By construction,
the only nonzero matrix element of $D^{i}$ is $D_{i}^{i}=1$. We
see that $E^j_{mn}=0$ as long as $m\ne n$ and $m$ and $n$ do not
equal $j$. Thus, matrix $E^{j}$ only has nonzero elements
of the form $E_{jm}^{j}=\bigl(E_{mj}^{j}\bigr)^*$ and $E_{mm}^{j}$, where $z^*$ denotes the complex conjugate of $z$. Note also by
setting $m=i$ and $n=j$ in Eq.~\eqref{eq: Vanishing E elements}
that $E_{ij}^{i}=-E_{ij}^{j}$ for $i\neq j$.

It remains to solve the last equation in \re{eq: Analogue}. Using
the above properties of matrix elements of $E^i$, we rewrite this
equation as
\begin{equation}
\begin{array}{l}
\dis
 E_{ii}^{n}-E_{mm}^{n}=\frac{E_{nm}^{m} E_{mn}^{m}}{E_{nn}^{m}-E_{ii}^{m}},\quad
i\neq m,n,\\
\\
\dis E_{ni}^{n}=\frac{E_{nm}^{m}E_{mi}^{m}}{E_{nn}^{m}-E_{ii}^{m}},\quad i\neq m,n,\\
\\
\dis E_{nn}^{n}-E_{mm}^{n}=E_{mm}^{m}-E_{nn}^{m}+\sum_{j\neq
m,n}\frac{E_{mj}^{m} E_{jm}^{m}}{E_{nn}^{m}-E_{jj}^{m}}.\\
\end{array}
\label{melE}
\end{equation}
By direct computation, one can show that   the following ansatz
  satisfies Eqs.~\eqref{melE}:
\begin{equation}
\begin{array}{l}
\dis E_{mj}^{m}=e^{\imath(\theta_{m}-\theta_{j})} \frac{\gamma_{m}\gamma_{j}}{\varepsilon_{m}-\varepsilon_{j}},\\
\\
\dis
 E_{jj}^{m}=-\frac{\gamma_{m}^{2}}{\varepsilon_{m}-\varepsilon_{j}}+\psi^{m},\quad
i\neq m,\\
\\
\dis E_{mm}^{m}=-\sum_{i\neq
m}\frac{\gamma_{i}^{2}}{\varepsilon_{m}-\varepsilon_{i}}+\psi^{m}.\\
\end{array}
\label{Ege}
\end{equation}
 where $\gamma_{j}\ne0$,  $\varepsilon_{j}$, $\theta_m$, and $\psi^{m}$   are real
parameters. A nonzero value of $\psi^m$ corresponds to an overall
shift of the diagonal of $E^m$, which yields a (nonessential)
contribution $\psi^{m}I$ to $h^m(u)$.
 Note
that $\gamma_{j}=0$ also satisfies Eqs.~\re{melE} but generates
matrices with block diagonal structure and, therefore,
$u$-independent symmetry.

Furthermore, any solution of Eqs.~\re{melE} admits parametrization
\re{Ege}. To establish this, it is sufficient to show that  any
choice of the $3N-2$ matrix elements $E_{mn}^{m}=\bigl(E_{nm}^{m}\bigr)^*$ and $E_{nn}^{m}$
for a certain $m$ compatible with Eqs.~\re{melE} corresponds to a
set of $3N+1$ real parameters, $\theta_{j}$, $\gamma_{j}$, $\varepsilon_{j}$, and $\psi^{m}$.
Then, Eqs.~\re{melE} ensure that all $E^j$ are of the form \re{Ege}.
The extra three parameters are an overall scale for
$\varepsilon_{j}$ and $\gamma_{j}$, a shift
$\varepsilon_{i}\to\varepsilon_{i}+\mbox{const}$, and a shift $\theta_{j}\to\theta_{j}+\mbox{const}$, which do not
affect Eqs.~\re{Ege}. To see the correspondence between the
$E_{mn}^{m}=\bigl(E_{nm}^{m}\bigr)^*$ and $E_{nn}^{m}$ and the
$\theta_{j},\gamma_{j},\varepsilon_{j},$ and $\psi^{m}$, note that Eqs.~\eqref{Ege}
yield
$$
 \sum_{n\ne
m}\dfrac{E_{mn}^{m} E_{nm}^{m}}{E_{nn}^{m}-\psi^{m}}=E_{mm}^{m}-\psi^{m},
$$
 which, for given $E_{mn}^{m}=\bigl(E_{nm}^{m}\bigr)^*$ and $E_{nn}^{m}$, can be solved for $\psi^{m}$. In seeking a common denominator, we see that it constitutes an $N^{\mathrm{th}}$ order polynomial
 \beg
 \prod_{j}{(E_{jj}^{m}-\psi^{m})}-\sum_{n\ne m}\prod_{j\ne m,n}{E_{mn}^{m} E_{nm}^{m}(E_{jj}^{m}-\psi^{m})}=0,
 \label{charact}
 \en
yielding $N$ solutions $\psi^{m}$. By considering the form of matrix $E^{m}$, i.e. that all matrix elements are zero save for a row, the corresponding column, and the diagonal, we find that the left hand side of \eref{charact} is the characteristic polynomial given by $\det{(E^{m}-\psi^{m} I)}$. Consequently, all $N$ solutions of \eref{charact} are guaranteed to be real as they are the eigenvalues of an explicitly Hermitian matrix. Once $\psi^{m}$ is determined, we can calculate ratios
$$
\gamma_{j}=-e^{\imath(\theta_{j}-\theta_{m})}\dfrac{E_{mj}^{m}}{E_{jj}^{m}-\psi^{m}}\qquad j\ne m,
$$
where $e^{2\imath(\theta_{j}-\theta_{m})}\equiv E_{jm}^{m}/E_{mj}^{m}$, $\theta_{m}$ is arbitrary and by a choice of an overall scale we  set $\gamma_{m}=1$. Lastly, letting $\varepsilon_{m}=0$ (by shifting
$\varepsilon_{i}$), we have
$$
\varepsilon_{j}=\dfrac{1}{E_{jj}^{m}-\psi^{m}}.
$$
Each of the $N$  solutions $\psi^{m}$ to \eref{charact} will yield a distinct set $\{\gamma_{j},\epsilon_{j}\}$, but by construction (see \eref{charact}) corresponds to the same set $\{E_{mn}^{m},E_{nm}^{m},E_{jj}^{m}\}$.

Now, consider $E^{m}$ as defined in \eref{Ege}. The matrix $E^{m}$ with complex matrix elements $E_{mj}^{m}=(E_{jm}^{m})^\ast$, $E_{jj}^{m}$, and $E_{mm}^{m}$ is conjugate to a matrix $\widetilde{E}^{m}$, i.e.
$$
E^{m}=\Sigma \widetilde{E}^{m} \Sigma^{-1},
$$
where $\Sigma$ is a diagonal matrix with entries $\Sigma_{jj} \equiv e^{\imath\theta_j}=e^{\imath\theta_m}\sqrt \frac{E_{jm}^{m}}{E_{mj}^{m}}$,  $(\widetilde{E}_{mj}^{m})^{2}=E_{mj}^{m} E_{jm}^{m}$, $\widetilde{E}_{jj}^{m}=E_{jj}^{m}$, and $\theta_{m}$ is an arbitrary real number. Given a Hermitian $E^{m}$, we find that $\widetilde{E}^{m}$ is necessarily real symmetric matrix and $\Sigma$ is a conjugating diagonal matrix whose matrix elements are complex phases. Thus, all Hermitian maximally commuting operators are matrix conjugate to some real symmetric such. Operator spectra are invariant under matrix conjugation and, therefore, it is convenient to henceforth limit our discussion to maximally commuting real symmetric matrices, and we do so without loss of generality.

Moreover, as noted below\eref{Ege},  nonzero $\psi^i$ contributes only a multiple of the
identity, $\psi^iI$ to each $h^i(u)$, which affects neither the
commutation relations nor the level crossings of $h^i(u)$ and
their linear combinations. Henceforth, we adopt a convenient
``gauge choice'' $\psi^{i}=0$ for all $i=1,\dots,N$ and $\theta_m=0$ for all $m=1,\dots,N$. With this
choice of $\psi^i$ and $\theta_m$, we derive from Eqs.~(\ref{canbas},\ref{Ege})
and the definition of $D^i$ the nonzero matrix elements of basic operators
$h^i(u)$
 \begin{equation}
\begin{array}{l}
\dis \left[h^{i}\left(u\right)\right]_{ij}  =
\frac{\gamma_{i}\gamma_{j}}{\varepsilon_{i}-\varepsilon_{j}},\quad i\ne j,\\
 \\
\dis \left[h^{i}\left(u\right)\right]_{jj} =
-\frac{\gamma_{i}^{2}}{\varepsilon_{i}-\varepsilon_{j}},\quad i\ne j,\\
\\
\dis\left[h^{i}\left(u\right)\right]_{ii}   =   u-\sum_{j\neq
i}\frac{\gamma_{j}^{2}}{\varepsilon_{i}-\varepsilon_{j}}.\\
\end{array}
\label{hmel}
 \end{equation}
 Note that
\begin{equation}
\sum_{i=1}^N h^{i}\left(u\right)=uI.
 \label{eq: sum to identity}
\end{equation}
Expressions \re{hmel} for matrix elements constitute a complete,
explicit solution of commutation relations \re{smalh} or
equivalently \re{alcom} for maximally commuting Hamiltonians. Different choices of parameters $\gamma_j$ and $\varepsilon_{j}$ (factoring out overall scale of $\gamma_j$ and $\varepsilon_{j}$, a total shift of all $\varepsilon_{j}$, and the ``gauge freedom'' discussed above) yield distinct
families of such Hamiltonians.

\eref{hmel} also determines matrix elements\cite{aside} of a general maximally
commuting  operator \re{arbh}
 \begin{equation}
\begin{array}{l}
\dis \left[H \left(u\right)\right]_{mn}= \gamma_{m}\gamma_{n}
\left(\frac{d_{m} -d_{n} }{\varepsilon_{m}-\varepsilon_{n}}\right),\quad m\ne n,\\
\\
\dis\left[H \left(u\right)\right]_{mm}=u\, d_{m} -\sum_{j\neq
m}\gamma_{j}^{2}
\left(\frac{d_{m} -d_{j} }{\varepsilon_{m}-\varepsilon_{j}}\right)\\
\end{array}
\label{Hmel}
\end{equation}

Let us also note that a convenient approach to producing nontrivial
solutions of Eq.~\eqref{comcur} was developed by Shastry in
Ref.~\onlinecite{Shastry}. Interestingly, these solutions turn out
to be essentially equivalent to the maximally commuting set
constructed in this section, see Appendix for details.

\section{Algebraic properties of maximal Hamiltonians and an upper bound on the number of level crossings}
\label{algebra}

The above parametrization makes transparent a beautiful property
of maximal Hamiltonians -- the product of two maximal
operators is itself the $u$-dependent sum of maximal operators.
This property, as we demonstrate in this section, allows one to
express  a  general maximal Hamiltonian
$\widetilde{H}(u)$ as a polynomial in another such Hamiltonian
$H(u)$. We employ this polynomial expansion to
determine the maximum number of level crossings in the eigenvalue
spectrum of $H(u)$.

First, we express the product of two basic maximally commuting
operators $h^i(u)$ and $h^j(u)$ in terms of $u$-dependent linear
combinations of $h^k(u)$.
 Using \eref{hmel}, one can show
that
\beg
\begin{array}{lcl}
\dis h^{i}\left(u\right)\cdot h^{j}\left(u\right) &  = &\dis \dfrac{\gamma_{j}^{2}}{\varepsilon_{i}-
\varepsilon_{j}}h^{i}\left(u\right)+
\dfrac{\gamma_{i}^{2}}{\varepsilon_{j}-\varepsilon_{i}}h^{j}\left(u\right),\quad i\neq j,\\
\\
\dis h^{i}\left(u\right)\cdot h^{i}\left(u\right) & = &\dis \sum_{k=1}^N\Biggl[
u-\sum_{m\neq k}\dfrac{\gamma_{m}^{2}}{\varepsilon_{k}-
\varepsilon_{m}}\Biggr]h^{k}\left(u\right).\\
\end{array}
\label{hh}
\en
 Now consider two general maximally commuting Hamiltonians \re{arbh}
\beg
H\left(u\right)=\sum_{k=1}^{N}d_{k}\,h^{k}\left(u\right),\quad
\widetilde{H}\left(u\right)=\sum_{k=1}^{N}\widetilde{d}_{k}\,h^{k}\left(u\right),
\label{2gen}
\en
where without loss of generality we dropped multiples of identity in \eref{arbh}.
 From \eref{hh} we derive
\begin{equation}
H\left(u\right)\cdot \widetilde{H} \left(u\right)=
\sum_{k=1}^{N}\Biggl[u\, d_{k} \widetilde{d}_{k} -\sum_{m\neq k}\frac{\gamma_{m}^{2}
\left(d_{k} -d_{m} \right)
 (\widetilde{d}_{k} - \widetilde{d}_{m})}{\varepsilon_{k}-\varepsilon_{m}}\Biggr]h^{k}\left(u\right).
\label{qr}
\end{equation}
 This quasi-ring structure -- so called because, while the sum of
maximal Hamiltonians is maximal, the product is a
$u$-dependent sum of such and, therefore, not generally linear in
$u$ and not strictly a maximal operator -- suggests a
means of representing an element of a commuting maximal family by
any other, see \eref{exp} below.

A typical maximal  Hamiltonian $H(u)$ can be degenerate
only at  discrete  values of $u$. Note that the only alternative to the discrete (possibly empty) set is
a permanent degeneracy -- when two eigenvalues of $H(u)$ coincide at all $u$ \cite{matr}. Permanent degeneracies do not
occur for a generic choice of $d_k$ in \eref{2gen}.  Indeed, recall that $d_k$ are the eigenvalues of $V$ (see
below \eref{arbh}).
Since the eigenvalues of $H(u)=T+uV$ tend to those of $uV$ for large $u$, the spectrum of $H(u)$ is not degenerate as long as the $d_k$ are distinct and $V$ is itself nondegenerate.

 Consider $H(u)$ at any  $u$ where it is nondegenerate.
Any element of
its commutant   -- the set of all real symmetric operators that commute with $H(u)$ --
can be expressed as a polynomial in $H(u)$ of the order $N-1$,
i.e.
\begin{equation}
\widetilde{H} \left(u\right)=\sum_{\alpha=0}^{N-1}P_{\alpha} \left(u\right)\,
H^{\alpha}\left(u\right),
\label{exp}
\end{equation}
where
$H^{\alpha}\left(u\right)\equiv\left[H\left(u\right)\right]^{\alpha}$
and, as we will see shortly, $P_{\alpha}\left(u\right)$ are
rational functions of $u$. To see that $\widetilde{H}(u)$ can be
indeed written in terms of powers of $H(u)$, consider \eref{exp}
in the common eigenbasis of commuting operators $H(u)$ and
$\widetilde{H}(u)$ at a given $u$. Since eigenvalues $\w_m$ of
$H(u)$ are $N$ distinct real numbers, one can always find a
polynomial $R_{N-1}(\w)=\sum_{\alpha=0}^{N-1}
P_\alpha\w^\alpha$ of order $N-1$ with $N$ real coefficients
$P_{\alpha}$ so that $R_{N-1}(\w_m)=\widetilde{\w}_m$, where $\widetilde{\w}_m$ are
the eigenvalues of $\widetilde{H}(u)$. Indeed, the
equations $R_{N-1}(\w_m)=\widetilde{\w}_m$  are linear in $P_\alpha$ with a nonzero determinant.

Next, we observe from Eqs.~\re{2gen} and \re{qr} that
\begin{equation}
H^{\alpha}\left(u\right)=\sum_{k=1}^N Q_{k}^{\alpha}\left(u\right)\,h^{k}\left(u\right),
\label{ha}
\end{equation}
 where for $\alpha\ge 1$ $Q_{k}^{\alpha}\left(u\right)$ is an $\alpha-1$ order polynomial
in $u$  determined by recursively applying Eq.~\eqref{qr} and $Q_k^0=1/u$ as follows from \eref{eq: sum to identity}. Plugging \eref{ha} into \eref{exp}
and using the second equation in \re{2gen}, we obtain
\begin{equation}
\sum_{k=1}^N\sum_{\alpha=0}^{N-1}P_{\alpha}(u)\, Q_{k}^{\alpha}(u)\,h^{k}(u)=
\sum_{k=1}^N\widetilde{d}_{k}\,h^{k}(u).
\label{eq: elaborate}
\end{equation}
Since $h^{k}\left(u\right)$ are linearly independent at any $u\ne0$, i.e.
$\sum_{k=1}^N f_{k}(u)\,h^{k}(u)=0$ if and only if $f_k(u)\equiv 0$ for all $u\ne 0$   \cite{note1}, Eq.~\eqref{eq: elaborate} becomes
\begin{equation}
\sum_{\alpha=0}^{N-1} P_{\alpha}\left(u\right)\, Q_{k}^{\alpha}\left(u\right)=\widetilde{d}_{k},\quad k=1,\dots, N.
\label{poly}
\end{equation}
Note that because $Q_{k}^{\alpha}\left(u\right)$ are rational functions in $u$, $P_\alpha(u)$ are also rational functions.

Because $\widetilde{H}(u)$ is arbitrary, Eqs.~\re{poly} should have solutions for $P_{\alpha}\left(u\right)$ for any
$\widetilde{d}_{k}$ as long as $H(u)$ is nondegenerate. On the other hand, solutions cease to exist if and only if
$\det[Q_{k}^{\alpha}(u)]=0$, where $Q_{k}^{\alpha}(u)$ is regarded as the $\alpha k^{\mathrm{th}}$
matrix element of an $N\times N$ matrix. Using the fact that $Q_{k}^{\alpha}\left(u\right)$ is a polynomial
in $u$ of degree $\alpha-1$ for $\alpha\ge 1$ and $Q_k^0=1/u$, one can show that
$\det[Q_{k}^{\alpha}(u)]=\mathcal{P}(u)/u$, where $\mathcal{P}(u)$ is a polynomial in $u$ of order $\sum_{m=0}^{N-2}m$. The real roots of the
equation $\mathcal{P}(u_\gamma)=0$ are the values of
$u=\{u_\gamma \}$ where $H(u)$ is degenerate.
Thus, the maximum possible number of level crossings in the eigenvalue
spectrum of an $N\times N$ maximally commuting Hamiltonian is
\beg
M_c^{\max}=\frac{(N-1)(N-2)}{2}.
\label{mmax}
\en
The polynomial $\mathcal{P}(u)$ is of real coefficients and, therefore,  its complex roots come in conjugate pairs.
Consequently,
the number of real roots of {\bf $\mathcal{P}(u)$} falls from the maximum $M_c^{\max}$ in decrements of two. This enforces a parity
such that
the number of real roots is odd for integers of the form $4m$, $4m+1$ and even for integers $4m+2$, $4m+3$, $m\in \mathbb{N}$. Ostensibly,
when real roots of {\bf $\mathcal{P}(u)$} are degenerate their number need not correspond to the number of distinct crossings.
In principle, a multiple real root
of $\mathcal{P}(u)$ could correspond to a single pairwise crossing. Numerically, however, we have observed that such
multiplicities
occur only when more than two levels cross simultaneously, i.e. at the same value of $u$.

\section{\label{sec: Central Spin Equiv.}Mapping  to the Gaudin magnets}

In this section, we  show that maximally commuting Hamiltonians
$h^i(u)$   are equivalent to
the Gaudin magnets,
\begin{equation}
\hat h^{i}\left(B\right)=-B\, \hat
s_i^z+\sum_{k=1}^{N}\lefteqn{\phantom{\sum}}'
\frac{\hat{\vec{s}}_{i}\cdot\hat{\vec{s}}_k}{\varepsilon_{i}-
\varepsilon_{k}}, \quad i=1,\dots,N, \label{gaud1}
\end{equation}
in the  next
to highest weight sector, $J^z=J^z_{\max}-1$, where $J^z$ is the
$z$ projection of the total spin, $\hat{\vec{s}}_{i}=\{\hat{s}^{x}_{i},\hat{s}^{y}_{i},\hat{s}^{z}_{i}\}$ is the $i^{\mathrm{th}}$ spin 3-vector of magnitude $s_i$ and $[\hat{s}^{\alpha}_{i},\hat{s}^{\beta}_{j}]=\epsilon_{\alpha \beta \gamma} \hat{s}^{\gamma}_{i}\delta_{ij}$. This mapping is very useful as
Gaudin magnets (central spin Hamiltonians) have been extensively
studied
\cite{Gaudin,Ushveridze,Sklyanin1,Sklyanin2,Amico,Delft,Dukelsky}.
For example, an exact solution for the eigenstates and eigenvalues
is available\cite{Gaudin,Ushveridze}. We employ it in subsequent
sections to obtain the spectra of maximally commuting Hamiltonians
and to analyze their level crossings. This mapping also implies
that all our conclusions regarding maximal Hamiltonians, e.g. the
presence and the number of level crossings, quasi-ring structure
\re{hh} etc. can be immediately transferred to the corresponding
sector of Gaudin magnets and their derivative models, such as the
reduced BCS model\cite{BCS,Richardson,integr}. At the same time,
other sectors of the Gaudin model as well as more general
models\cite{Delft} of which it is a particular case can provide
examples of Hamiltonians with less then the maximum number of
commuting partners.

Since Gaudin magnets \re{gaud1} commute with the $z$ projection of the total spin $\hat J^z$, see \eref{Jz}, they are
block-diagonal in any basis where  $\hat J^z$ has a definite value. Different blocks can be labeled by the
eigenvalues of  $\hat J^z$. Consider the sector $J^z=J^z_{\max}-1$, where $J^z_{\max}=\sum_{i=1}^N s_i$ is
the maximum eigenvalue of $\hat J^z$. It is populated by $N$  basic states
\beg
|k\rangle
= \frac{\hat s_{k}^{-}|0\rangle}{\sqrt{2s_{k}}},\quad k=1,\dots,N,
\label{basic}
\en
where $|0\rangle$ is the highest weight state $J^z=J^z_{\max}$, i.e. $\hat s_k^+|0\rangle=0$ for all $k$, and the highest weight $s_{k}$ for each spin $\hat{s}_k$ is given by $\hat{s}_{k}^{z} |0\rangle=s_{k} |0\rangle$.
Therefore, Gaudin Hamiltonians  \re{gaud1} are $N$  commuting real symmetric
$N\times N$ matrices in this sector. Since there is also no obvious  $B$-independent symmetry
($\hat J^z\propto I$ within a given sector), the $\hat h^i(B)$ appear to be good candidates for a maximally
commuting set.

To check this, let us evaluate the nonvanishing matrix elements of $\hat h^i(B)$ given by \eref{gaud1} in the normalized
basis \re{basic}. We obtain
\begin{equation}
\begin{array}{l}
\dis \langle i |\hat h^{i} (B) |j \rangle
=\frac{\sqrt{s_{i}s_{j}}}{\varepsilon_{i}-\varepsilon_{j}},\quad j\ne i,\\
 \\
\dis  \langle j| \hat h^{i}\left(B\right)|j \rangle
=-\frac{s_{i}}{\varepsilon_{i}-\varepsilon_{j}}+\Biggl[-B\, s_{i}+\sum_{k\neq i}\frac{s_{i}s_{k}}{\varepsilon_{i}-
\varepsilon_{k}}\Biggr],\quad j\ne i,\\
\\
\dis  \langle i|\hat h^{i}(B)|i\rangle=B-\sum_{k\neq i}\frac{s_{k}}{\varepsilon_{i}-\varepsilon_{k}}+\Biggl[-B\, s_{i}+
\sum_{k\neq i}\frac{s_{i}s_{k}}{\varepsilon_{i}-\varepsilon_{k}}\Biggr].\\
\end{array}
\label{gaudmel}
\end{equation}
Comparing these expressions to matrix elements of $h^i(u)$ in \eref{hmel}, we observe
that with the identifications $B=u$ and $s_k=\gamma_k^2$ the two matrices differ only by a multiple of
an identity matrix $\psi^i I$, where
\beg
\psi^{i}= -B\, s_{i}+\sum_{k\neq i}\frac{s_{i}s_{k}}{\varepsilon_{i}-\varepsilon_{k}}.
\label{psi}
\en
Recall that we arbitrarily  selected a ``gauge'' $\psi^i=0$ for maximally commuting Hamiltonians
$h^i(u)$, see \eref{Ege} and the text above \eref{hmel}. This constant overall shift of
all eigenvalues of $h^i(u)$  affects neither its eigenstates nor the degeneracies.

Thus, we see that  Gaudin Hamiltonians \re{gaud1} in the next to highest weight sector
$J^z=J^z_{\max}-1$ are equivalent to basic maximal Hamiltonians $h^i(u)$ with
\beg
u=B,\quad \gamma_k^2=s_k,
\label{transl}
\en
and vice versa. Note that  the magnitudes of quantum spins, $s_{k}$, take half-integer values for finite
dimensional representations
of the spin $su\left(2\right)$ algebras, while $\gamma_k$ are
arbitrary real numbers. We believe that this restriction can be lifted
by moving to an appropriate infinite dimensional representations of the $su\left(2\right)$s, where
the highest weight states are still well defined but $s_k$ take arbitrary real values\cite{cruz}.
Indeed, we have  verified that, at least in our sector $J^z=J^z_{\max}-1$,
 in all expressions for the eigenvalues and eigenstates of $\hat h^i(B)$ (see below) the replacements $B\to u$ and
 $s_k\to\gamma_k^2$ with arbitrary real $\gamma_k$
produce the correct corresponding eigenvalues and eigenstates of $h^i(u)$.

\section{\label{sec: Exact Solution}Exact solution for the spectra of maximal Hamiltonians}

A particularly useful consequence of the mapping  \re{transl} between
Gaudin magnets $\hat h^i(B)$ and maximally commuting Hamiltonians $h^i(u)$ is that one can
obtain the exact solution for $h^i(u)$  by importing the
known exact solution for the spectra of  $\hat h^i(B)$ \cite{Gaudin,Ushveridze}.
The latter has been derived both  from the properties
of the  Gaudin algebra\cite{Gaudin} and  by Bethe's Ansatz \cite{Delft}.

The exact eigenvalues  of the Gaudin Hamiltonian \re{gaud1}, $\hat h^i(B)$, in the
next to highest weight sector $J^z=J^z_{\max}-1$ are
\beg
\left(\lam_m^i\right)_G=\frac{s_i}{x_m^G-\eps_i} +\psi^i,
\label{egaud}
\en
where $\psi^i$ is the overall shift of all eigenvalues given by \eref{psi} and $x_m^G$ are
the solutions of the following equation:
\beg
B=\sum_{k=1}^N\frac{s_j}{x_m^G-\eps_k}.
\label{xmg}
\en
Note that if this equation is brought to the common denominator, the numerator becomes a polynomial
of order $N$ in $x_m^G$. Therefore, there are $N$ solutions for $x_m^G$ and $N$  eigenvalues \re{egaud} as
it should be since there are $N$ states in this sector, see \eref{basic}. The unnormalized eigenstates
(common to all $\hat h^i(B)$) corresponding to eigenvalues \re{egaud} are
\beg
|\lam_m\rangle\lefteqn{\phantom{\bigr)}}_G=\sum_{k=1}^N\frac{\sqrt{s_k} |k\rangle}{x_m^G-\eps_k},
\label{sgaud}
\en
where the basic states $|k\rangle$ have been introduced in \eref{basic}. A concise derivation of
Eqs.~(\ref{egaud},\ref{xm}) and \re{sgaud} can be found in Refs.~\onlinecite{Gaudin}, \onlinecite{Ushveridze},
and \onlinecite{Dukelsky}.

Using the mapping \re{transl} between basic maximal operators $h^i(u)$ and Gaudin Hamiltonians, we
obtain from \eref{egaud} the energies of $h^i(u)$
\beg
 \lam_m^i=\frac{\gamma^2_i}{x_m -\eps_i}.
\label{e}
\en
Note that we set the overall shift $\psi^i=0$ in accordance to the discussion surrounding \eref{psi}.
The corresponding common eigenstates of all $h^k(u)$ are
\beg
|\lam_m\rangle =\sum_{k=1}^N\frac{ \gamma_k |k\rangle}{x_m-\eps_k},
\label{s}
\en
where $|k\rangle$ now stands for a basic vector for matrices
$h^i(u)$, i.e. its $j^{\mathrm{th}}$ component is $|k\rangle_j=\delta_{jk}$.
In Eqs.~\re{e} and \re{sgaud} $x_m$ are solutions of the following equation:
\begin{equation}
u=\sum_{k=1}^N\frac{\gamma_{k}^{2}}{x_{m}-\varepsilon_{k}}\equiv f(x_m),\quad m=1,\dots,N,
\label{xm}
\end{equation}
which follows from \eref{xmg}. That Eqs.~(\ref{e},\ref{s}) and \re{xm} yield the correct spectrum of
$h^i(u)$ can be verified directly using the matrix form \re{hmel} of $h^i(u)$. Finally, using
\eref{2gen}, we derive the energies of a general maximally commuting Hamiltonian, $H(u)=T+uV$,
\begin{equation}
\omega_{m}=\sum_{k=1}^{N} \frac{d_{k}\gamma_{k}^{2}}{ x_m-\varepsilon_{k}}.
\label{eH}
\end{equation}
The corresponding eigenstates are still given by \eref{s}.

\begin{figure}
\begin{centering}
\includegraphics[scale=0.8]{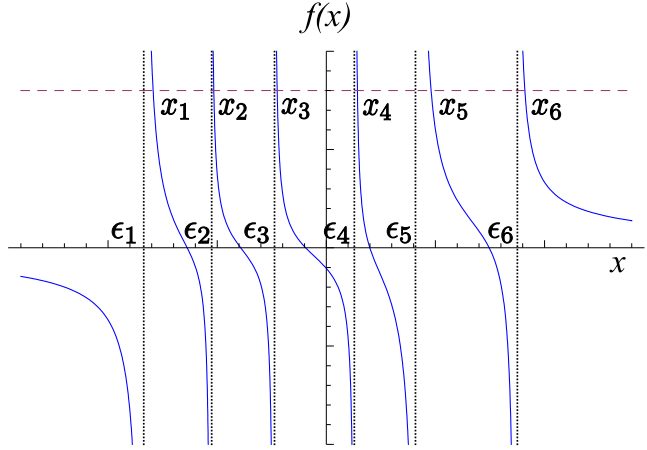}
\end{centering}
\caption{A plot of $f(x)=\protect\sum_{k=1}^N \gamma_k^2(x-\eps_k)^{-1}$ for $N=6$. Solutions of the equation $f(x_m)=u$ determine
the energies \re{eH} and eigenstates \re{s} of a general maximal Hamiltonian $H(u)$.
There are $N$ points, $x_m$ with $m=1,\dots,N$, where $y=f(x)$ intersects $y=u$ (dashed horizontal line) yielding $N=6$ eigenstates.
Note that $\eps_m<x_m<\eps_{m+1}$ except in the case of $x_N$ for which we have $\eps_N<x_N$ for $u>0$ and
$x_N<\eps_{N+1}\equiv \eps_1$ for $u<0$. Furthermore, we see that $x_m\to\eps_m$ as $u\to\infty$ and
$x_m\to\eps_{m+1}$ as $u\to-\infty$. This observation allows us to determine the behavior of the energies in the $u\to\pm\infty$ limits,
see \eref{hpr}.  }
\label{branches}
\end{figure}

Let us analyze the flow of eigenvalues $\w_m$ of $H(u)$ with $u$ and determine their
behavior in the $u\to\pm\infty$ limits. Consider \eref{xm}. Let $\eps_k$ be ordered as $\eps_1<\eps_2<\dots<\eps_N$.
The right hand side of \eref{xm} is plotted in Fig.~\ref{branches}. Note that $f(x)\to +\infty$ as $x\to \eps_k^+$
and $f(x)\to -\infty$ as $x\to \eps_{k+1}^-$. It follows that the equation $u=f(x_m)$ has a real root between $\eps_k$ and
$\eps_{k+1}$ for any $k$. Let us number the roots $x_m$ so that $\eps_m<x_m<\eps_{m+1}$. Note from Fig.~\ref{branches}
that for the last root $x_N$ we have $\eps_N<x_N$ for $u>0$ and $x_N<\eps_{N+1}\equiv \eps_1$ for $u<0$, where from
now on we identify
indices $m$ and $m+N$ that differ by a multiple of $N$. Further, observe  from Fig.~\ref{branches}  that
$x_m\to \eps_m$ as $u\to+\infty$. In this limit $k=m$ term dominates  Eqs.~\re{xm} and \re{eH} and
we obtain $\gamma_m^2/(x_m-\eps_m)\approx u$ and $\w_m\approx d_m\gamma_m^2/(x_m-\varepsilon_{m})\approx u d_m$.
Similarly, for $u\to-\infty$ we have $x_m\to \eps_{m+1}$ and $\w_m\approx ud_{m+1}$. Therefore,
\beg
\w_m\to -|u| d_{m+1}\quad\mbox{as $u\to-\infty$},\quad \w_m\to |u| d_{m}\quad\mbox{as $u\to+\infty$}.
\label{hinf}
\en

At this point it is convenient to rescale the Hamiltonian
\beg
H'(u)=\frac{H(u)}{\sqrt{u^2+1}}=\frac{T+uV}{\sqrt{u^2+1}}.
\label{scale}
\en
Note that this does not affect the level crossings, i.e. $H'(u)$ and $H(u)$ have crossings (if any) at the same values
of $u$. \eref{hinf} implies
\beg
\w'_m\to -d_{m+1}\quad\mbox{as $u\to-\infty$},\quad \w'_m\to d_{m}\quad\mbox{as $u\to+\infty$},
\label{hpr}
\en
where $\w'_m$ is the eigenvalue of $H'(u)$ corresponding to the eigenstate $|\lam_m\rangle$. Recall that $d_k$ are
the eigenvalues of $V$, see the text below \eref{arbh}. We see from \eref{scale} that the eigenvalues of $H'(u)$
indeed should tend to $d_k$ in  $u\to\pm\infty$ limits consistent with \eref{hpr}. The latter equation however provides
much more detailed information -- it shows to which particular $d_k$ the eigenvalue corresponding to a given eigenvector
tends in each limit. We will use \eref{hpr} in the next section to study the crossings of energy levels of a general
maximally commuting Hamiltonian $H(u)$.

\section{\label{sec: Crossings} Level crossings}

In this section, we establish the presence of energy level
crossings in the spectrum of an arbitrary maximally commuting
Hamiltonian $H(u)=T+uV$ \re{arbH}. This provides an explanation of
the level crossing phenomenon in the absence of any
$u$-independent symmetry based  solely on the fact that $H(u)$ has
the maximum possible number of independent commuting partners, see
the text above \eref{alcom}. Further, we determine the number of
level crossings as it depends on the ordering of the eigenvalues $d_k$
of the perturbation operator $V$ and argue that this number takes
values \beg M_c=\frac{(N-1)(N-2)}{2} - 2K,\quad K=0,1,\dots,
K_{\max}, \label{crossnum} \en where $N$ is the dimensionality of
the state space of $H(u)$ and $K_{\max}$ is the integer part (floor) of
$(N-1)(N-2)/4-1/2$. For example, $N=3$ maximally commuting
operators have a single level crossing, while for $N=6$ we have
$M_c=2,4,6,8,$ and 10. We also develop an approach that allows   us to
  readily predict the  minimum allowed number of crossings for a given $H(u)$
from the ordering of $d_k$.

Consider a Hamiltonian (not necessarily belonging to any commuting family)  that depends on a real parameter $u$.
Suppose $|n_i\rangle$ are its eigenstates and $E^-_{n_1}<E^-_{n_2}<\dots$ are the corresponding
energies  at large negative $u$. There is only one way to avoid crossings -- the order of
eigenvalues $E^+_{n_i}$ at $u\to\infty$ must be  exactly the same as that at $u\to-\infty$, i.e. $E^+_{n_1}<E^+_{n_2}<\dots$
This is what   happens with a typical Hamiltonian in agreement with the Wigner-von Neumann non-crossing rule,
Fig.~\ref{fig: avoided crossings}. If, on the
other hand, the relative order of any two energies changes, at least one level crossing must occur. For example,
$E^-_{n_1}<E^-_{n_4}$ and $E^+_{n_1}>E^+_{n_4}$ means that the difference $E_{n_1}(u)-E_{n_4}(u)$  changes sign as
$u$ evolves from $-\infty$ to $\infty$. By continuity this implies a crossing of levels corresponding
to eigenstates $|n_1\rangle$ and $|n_4\rangle$ at a certain value of $u$. This
 is observed in blocks of quantum integrable Hamiltonians characterized by the same $u$-independent symmetry,
see e.g. Fig.~\ref{fig: typical integrable}. Numerical spectra of maximal Hamiltonians display the same
behavior, Figs.~\ref{real5} and \ref{fig: double cross}a.

\begin{figure}[t!]
\begin{centering}
\includegraphics[width=0.4\textwidth]{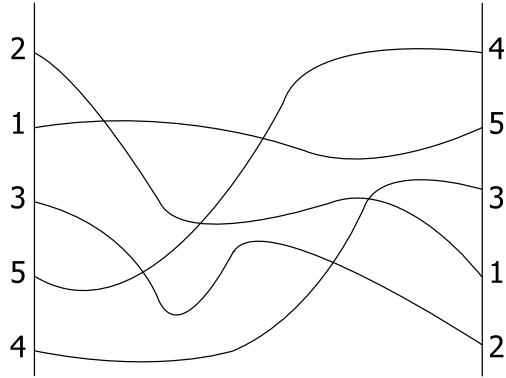}
\end{centering}
\caption{\label{schem5} Schematic energy level diagram for general $5\times5$ maximal Hamiltonians $H(u)=T+uV$ drawn using
the rules \re{rule}. The eigenvalues $d_{k}$ of $V$ are ordered such that
$d_{4}>d_{5} >d_{3} >d_{1} >d_{2}$. The diagram shows six level crossings for this ordering and specifies which levels cross,
e.g. the level $4\to3$ first crosses $3\to2$ and then $2\to1$. Compare to Fig.~\ref{real5} and note that the crossings
predicted by the above diagram are exactly the same as actual numerical crossings for this ordering. Note also that according
to \eref{crossnum} six is the maximum allowed number of crossings for this ordering of $d_k$
and multiple crossings of the same two levels are therefore forbidden for this ordering.  }
\end{figure}

\begin{figure}[p]
\begin{centering}
\includegraphics[scale=0.8]{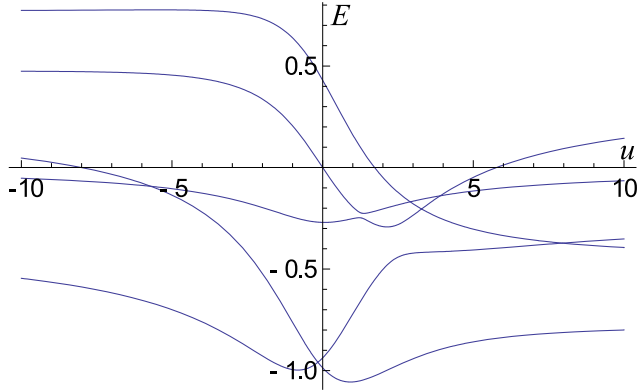} %\includegraphics[scale=0.8]{realschematic.pdf}
\end{centering}
\caption{\label{real5} Numerical energy levels of a  $5\times5$ maximal Hamiltonian $H(u)=T+uV$ with the same ordering
of eigenvalues of $V$ as that in Fig.~\ref{schem5}. Energies are scaled with a factor $(u^{2}+1)^{-1/2}$ to highlight their
asymptotic approach  to eigenvalues of $V$.
Matrix elements of $H(u)$ are generated using \eref{Hmel} with random
$\gamma_k$, $\eps_k$ and random $d_k$ constrained to obey the ordering of Fig.~\ref{schem5}.
Note that the number of crossings as well as the levels that cross are exactly the same as those predicted by Fig.~\ref{schem5}.}
\end{figure}

\begin{figure}[p]
\begin{centering}
\includegraphics[scale=0.3]{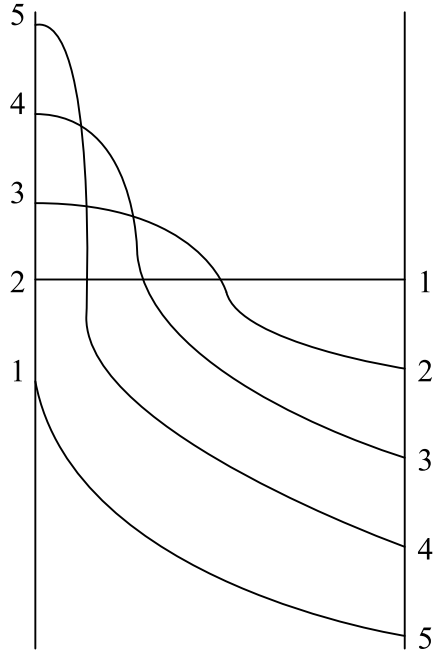} %\includegraphics[scale=0.8]{realschematic.pdf}
\end{centering}
\caption{\label{maxcross}  A schematic diagram corresponding to maximum level crossings with $N=5$.
As is evident,  $N\to N-1$ line has $N-2$ crossings, $N-1 \to N-2$ adds $N-3$ new crossings and so on, till
the line  $3\to2$  adds only 1 new crossing. Thus, the maximum number of crossings, $1+2+3+...+(N-2)= (N-1)(N-2)/2$,
can be confirmed.}
\end{figure}

\begin{figure}[h]
\begin{centering}
a)\includegraphics[width=0.2\textwidth]{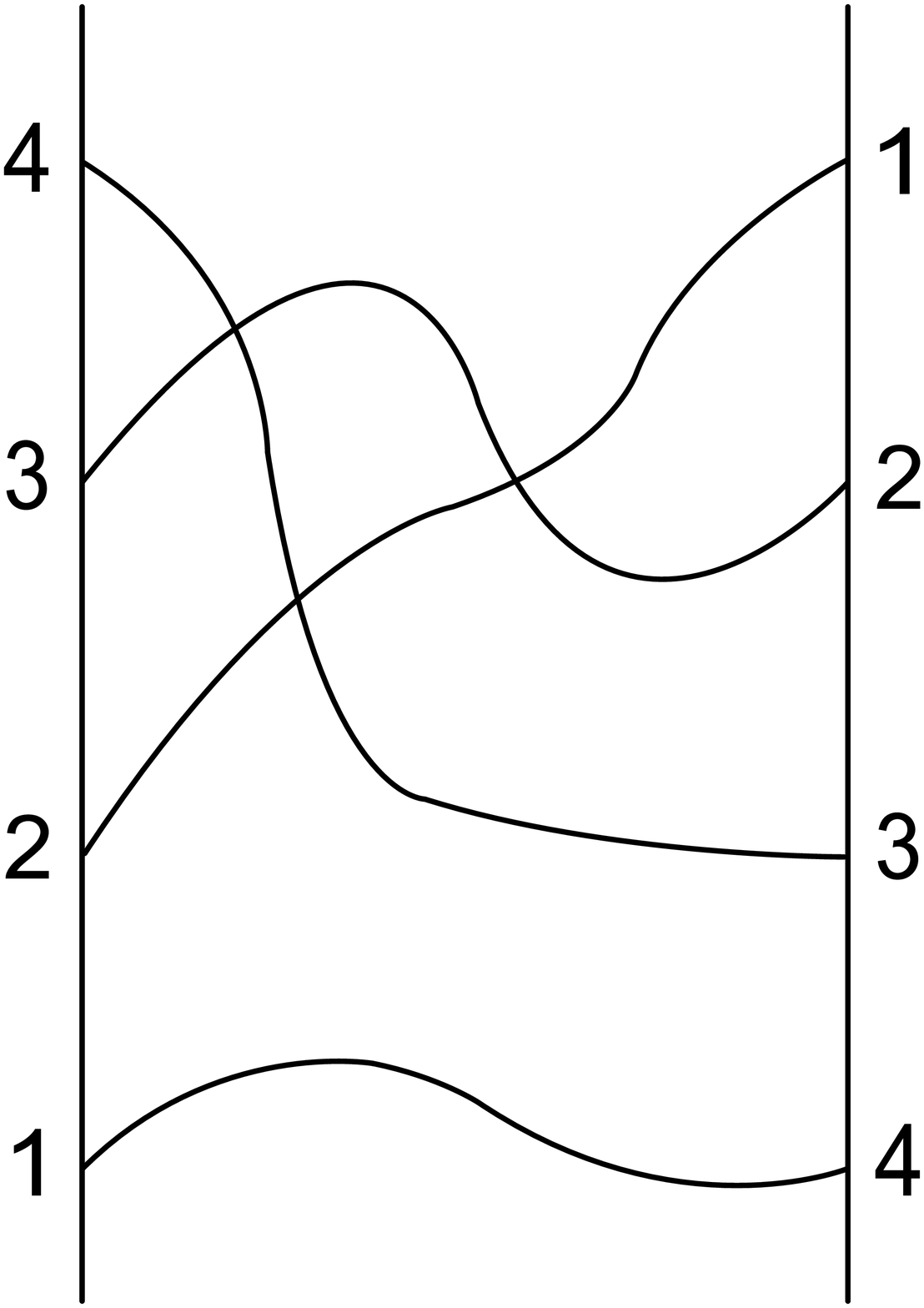}$\qquad$
b)\includegraphics[width=0.2\textwidth]{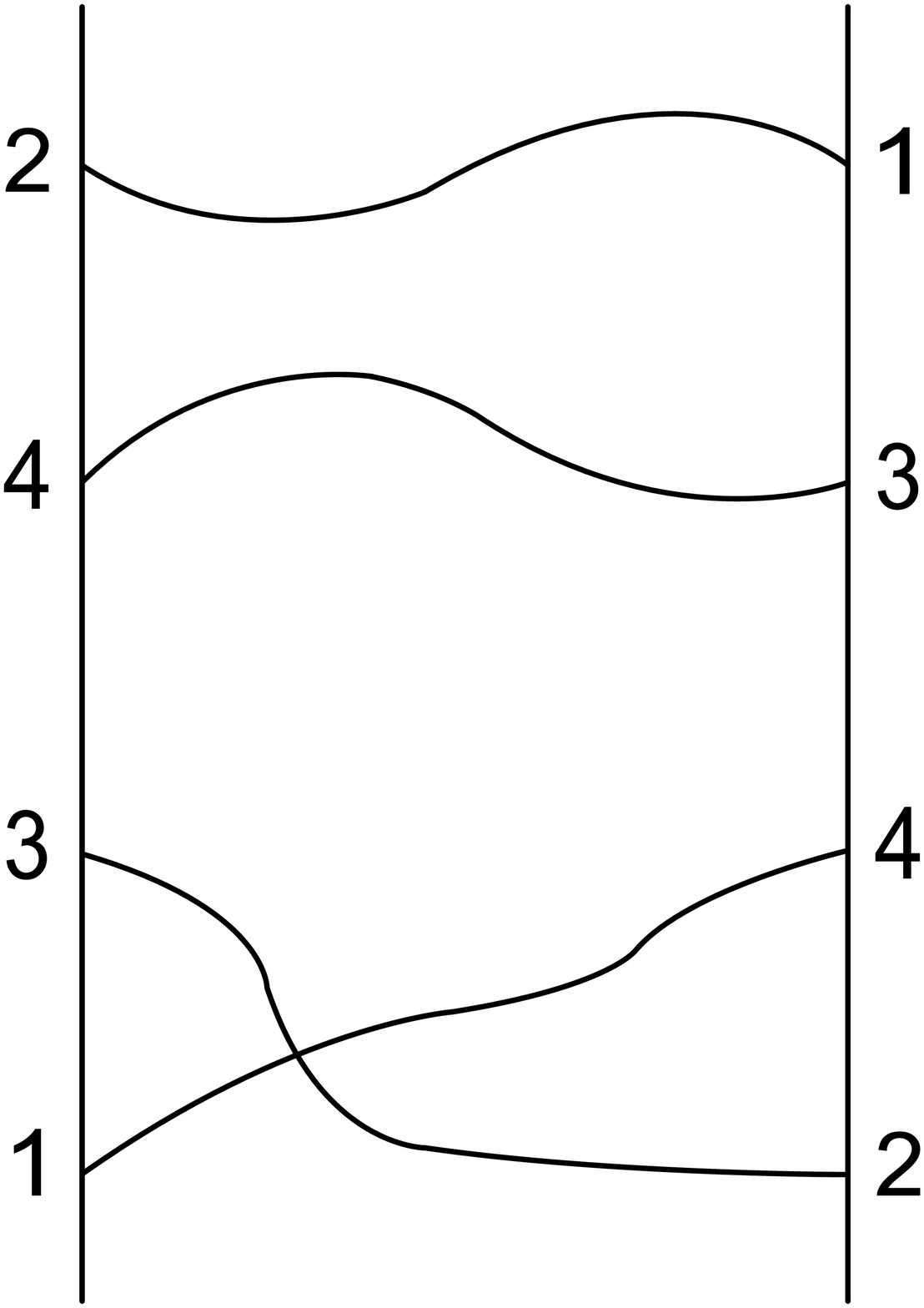}$\qquad$
c)\includegraphics[width=0.2\textwidth]{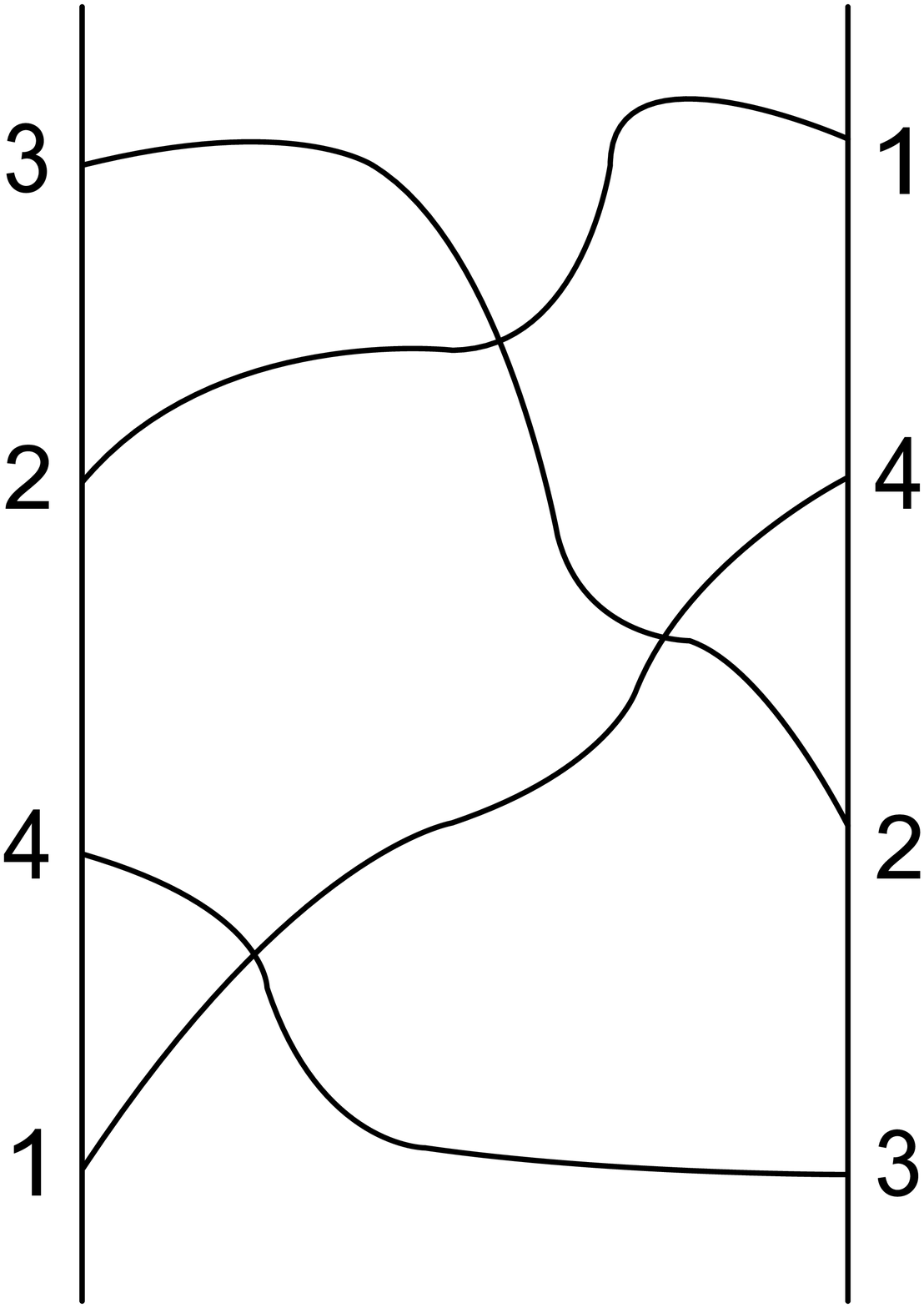}
\end{centering}

\medskip{}

\begin{centering}
d)\includegraphics[width=0.2\textwidth]{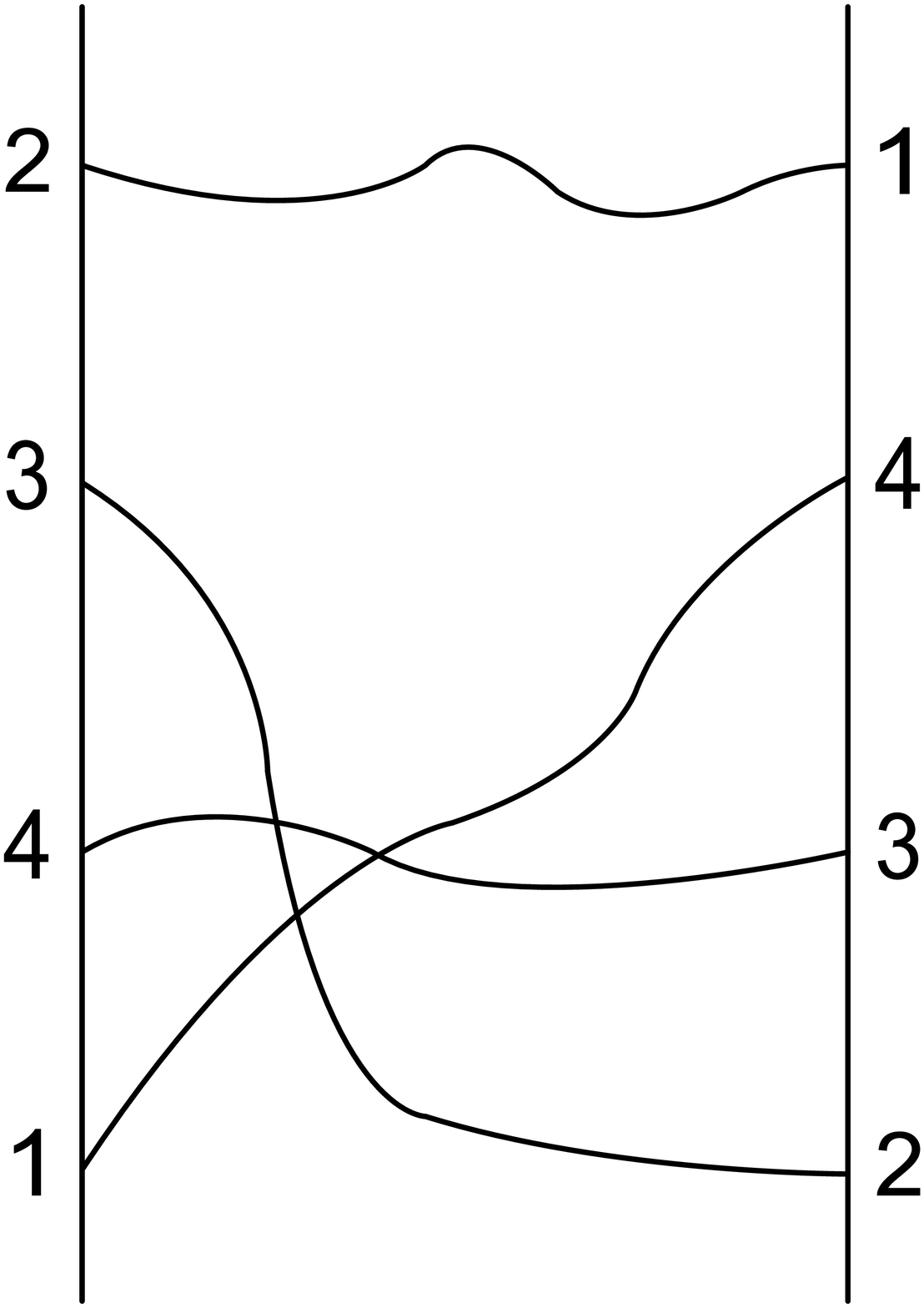}$\qquad$
e)\includegraphics[width=0.2\textwidth]{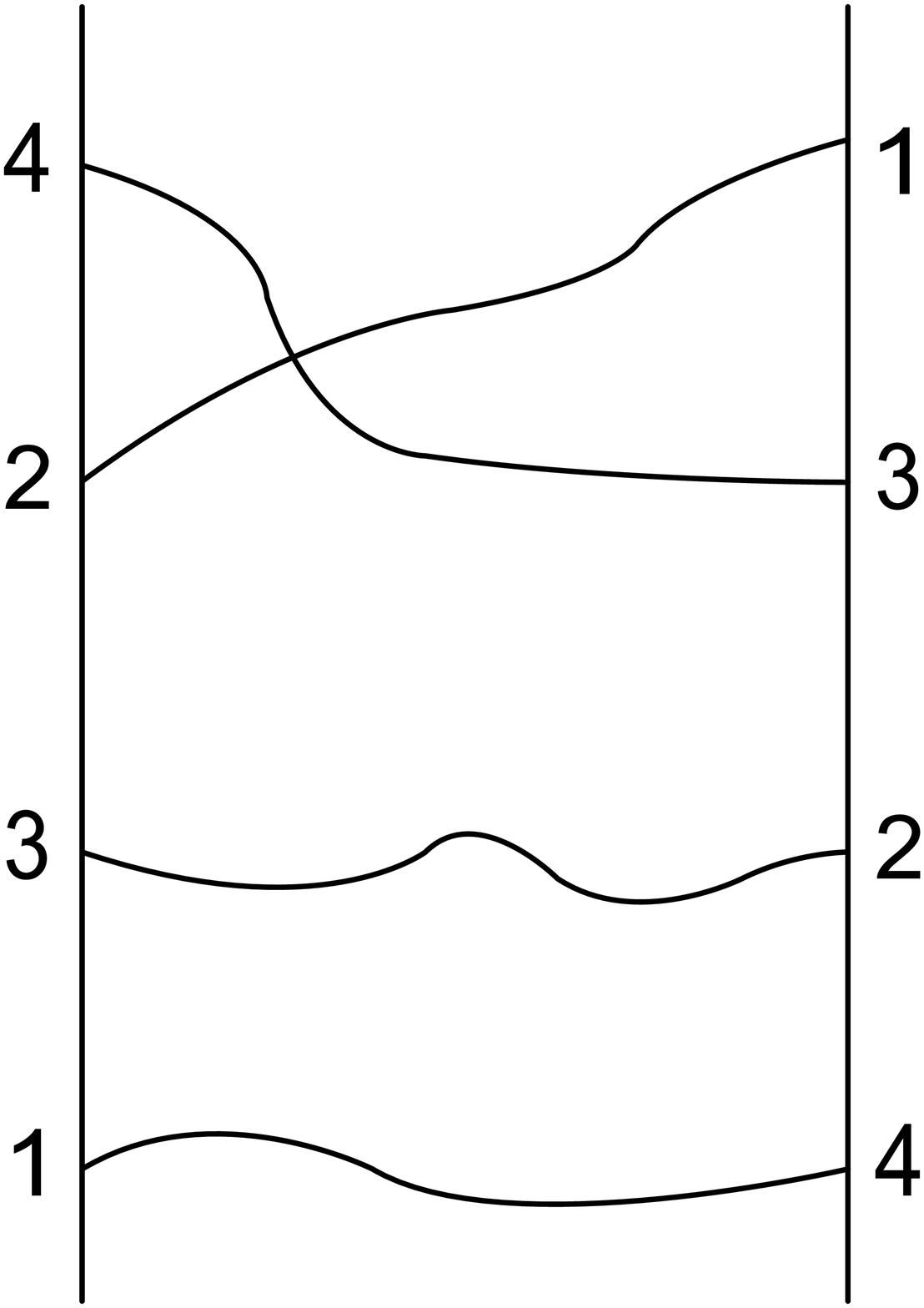}$\qquad$
f) \includegraphics[width=0.2\textwidth]{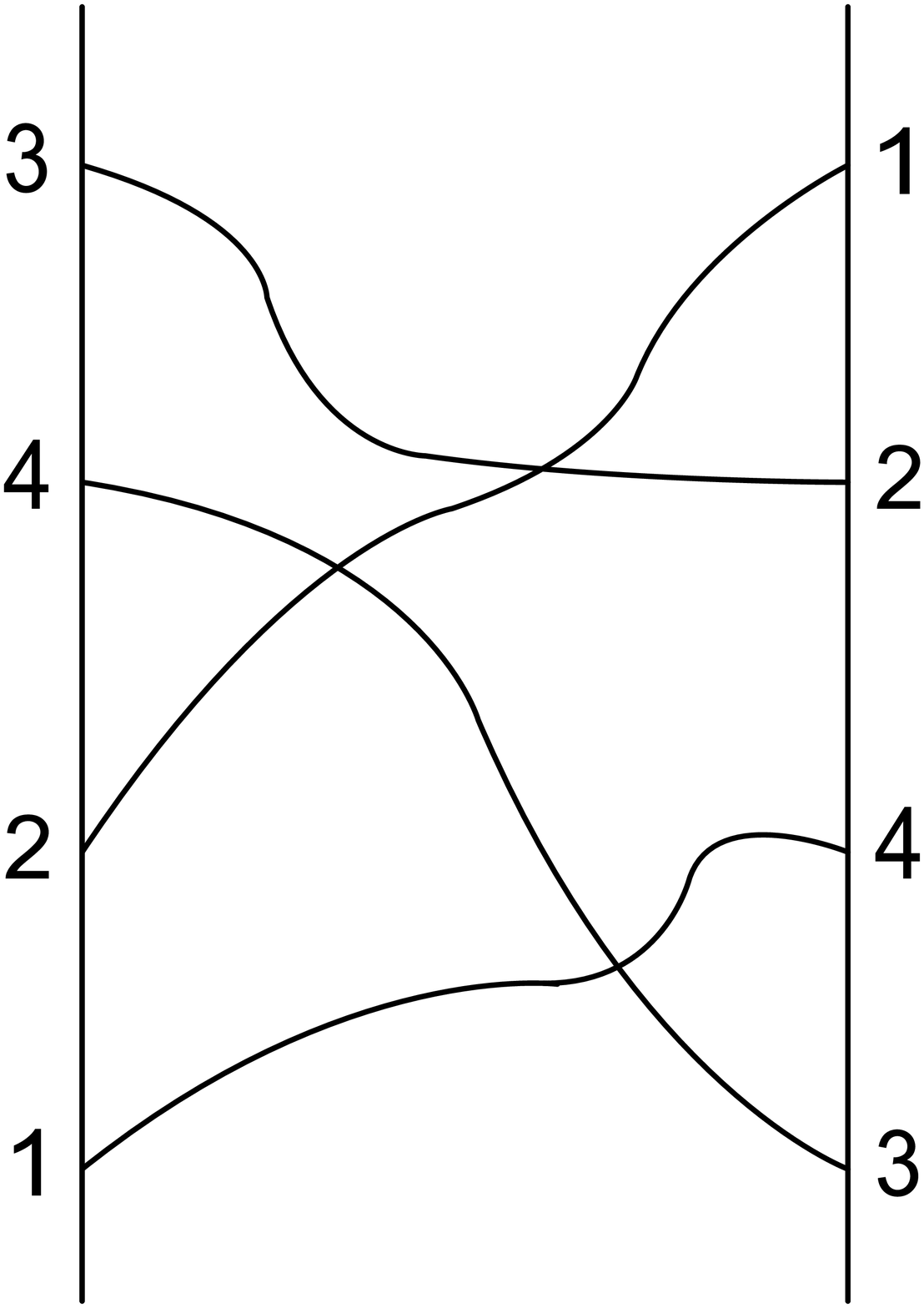}
\end{centering}

\caption{\label{4x4}  All distinct  level diagrams of $N=4$
maximal Hamiltonians, $H(u)=T+uV$, drawn according to \re{rule}. There are $(N-1)!=6$ distinct diagrams each corresponding
to $N=4$ different orderings of eigenvalues of   $V$ (see the text). For example, diagram b)
corresponds to $d_{1} >d_{3} >d_{4} >d_{2}$ and three other orderings obtained with a shift of the indices  by an integer
mod $N$, e.g. $d_{2} >d_{4} >d_{1} >d_{3}$ etc.
 The diagrams
  predict   either one or three level crossings in agreement with \eref{crossnum} and specify which levels cross. However,
  when the number of crossings is less than the maximum (three), additional multiple crossings of the same two levels
  can occur. This can increase the  number of crossings by $2K$, see Fig.~\ref{fig: double cross}. In the present case, the number
  of crossings for orderings b) and e) can increase from one to three. }
\end{figure}

\begin{figure}[p]
\begin{centering}
a)\includegraphics[height=4.6cm]{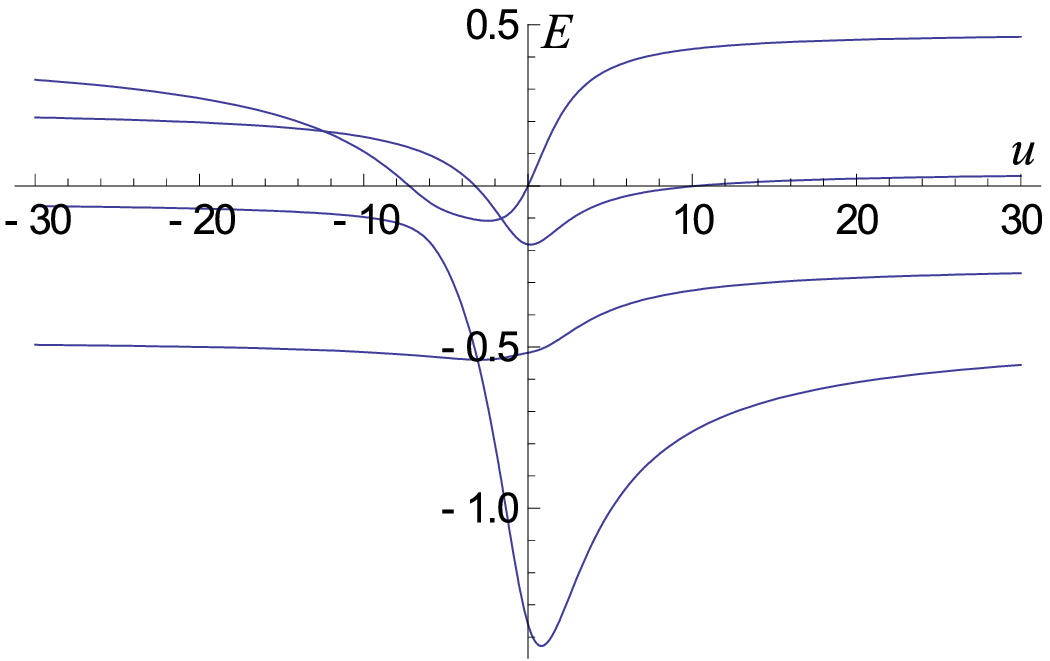}
\hspace{0.5cm}
b)\includegraphics[height=4.6cm]{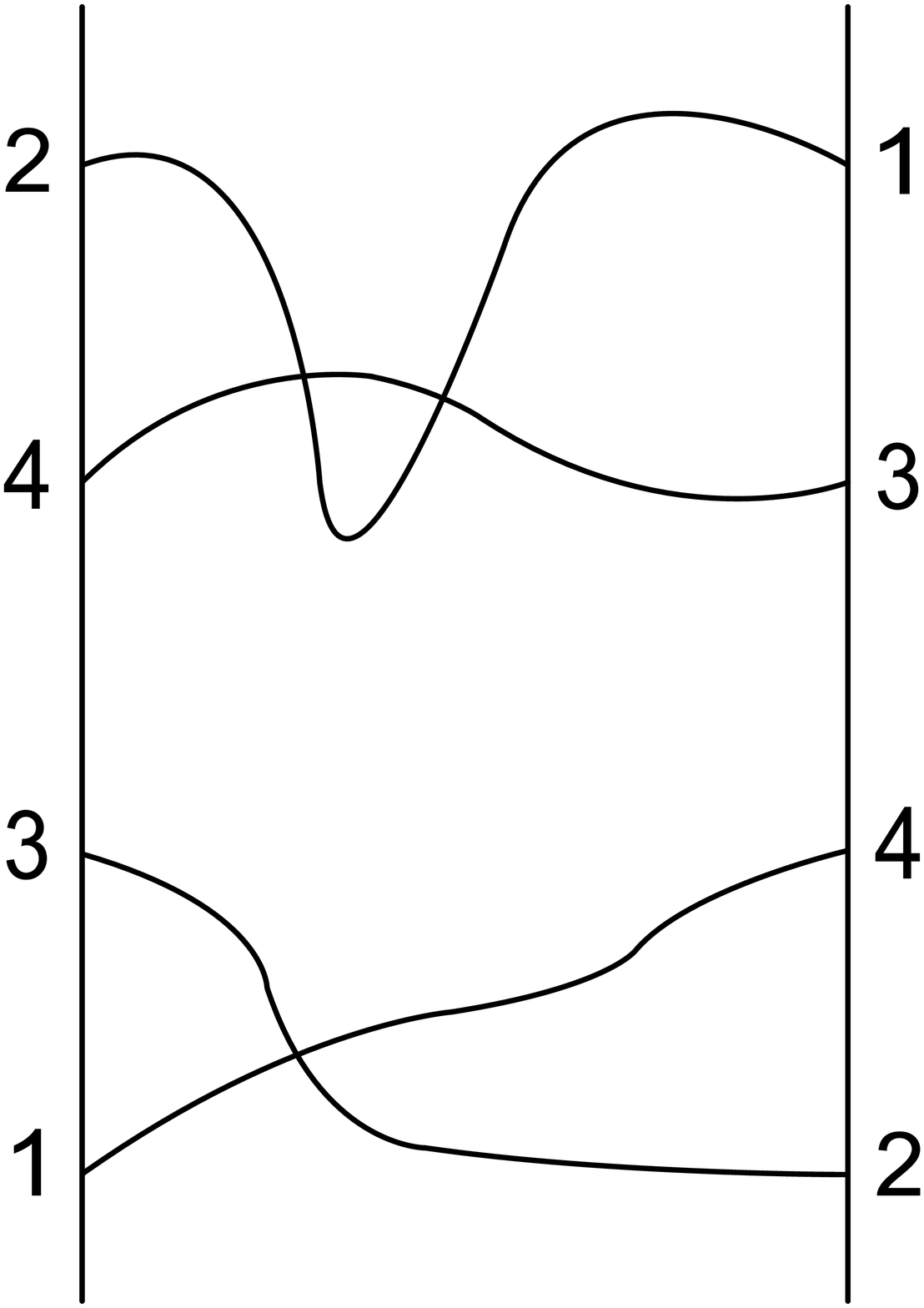}
\end{centering}
\caption{\label{fig: double cross}
  a) Numerical energy levels of a  $4\times4$ maximal Hamiltonian $H(u)=T+uV$ with the same ordering
of $d_k$ as  in Fig.~\ref{4x4}b. Energies are scaled with  $(u^{2}+1)^{-1/2}$ to highlight their
asymptotic approach  to $d_k$. Matrix elements of $H(u)$ are generated using \eref{Hmel} with random
$\gamma_k$, $\eps_k$ and random $d_k$ constrained to obey the above ordering.
 The multiple (twofold) crossing of the top two levels increases the number of crossings from one as in Fig.~\ref{4x4}b
 to three. This illustrates the generic situation arising when multiple crossings  increase the number of crossings by an
 even integer over and above  the number enforced by the diagrams \re{rule}. Nevertheless, as discussed in the text,
 this does not affect \eref{crossnum}.   b) The schematic of a).}
\end{figure}

\begin{figure}[h]
\begin{centering}
a)\includegraphics[width=0.25\textwidth]{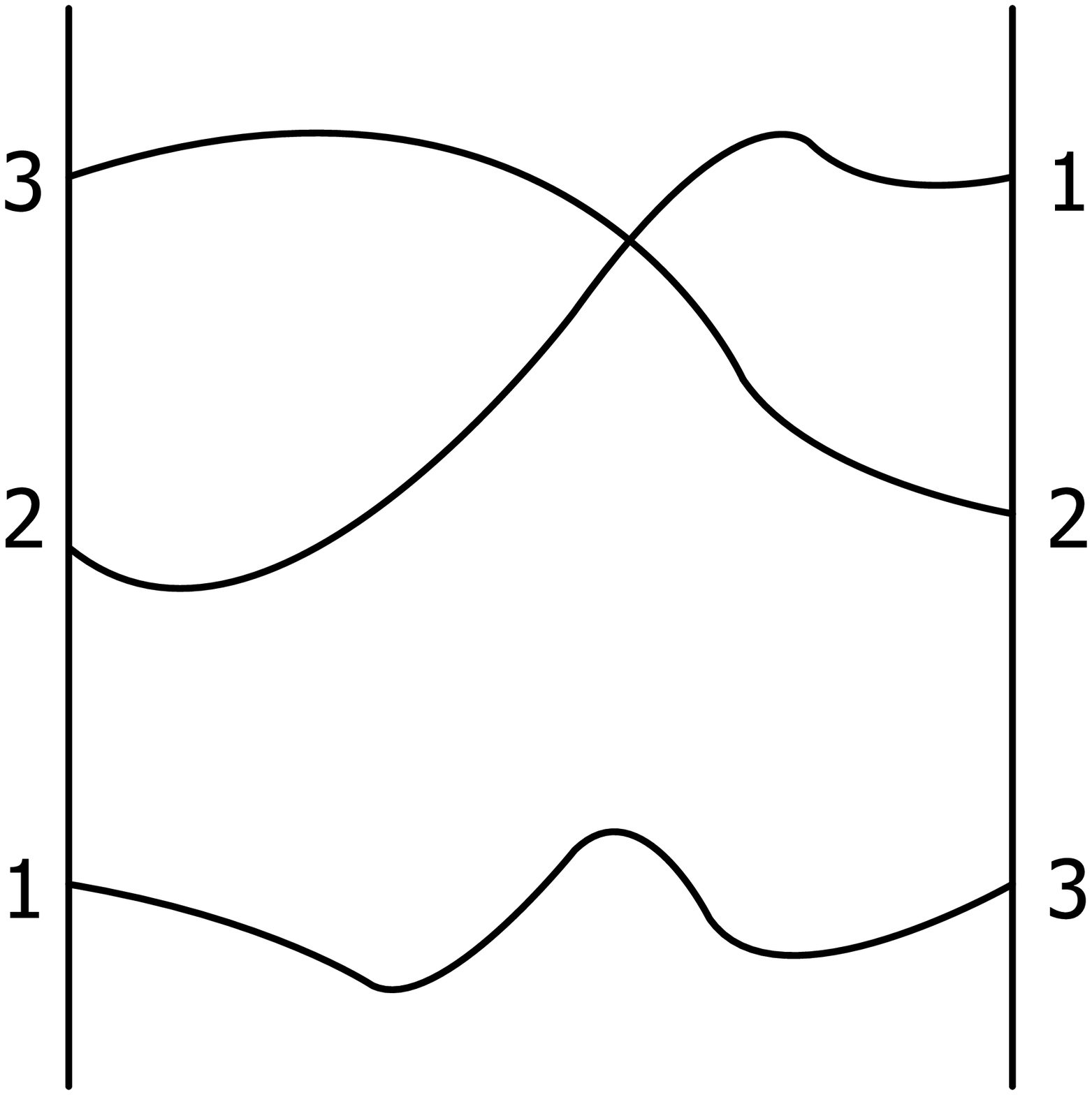}$\qquad\qquad$
b)\includegraphics[width=0.25\textwidth]{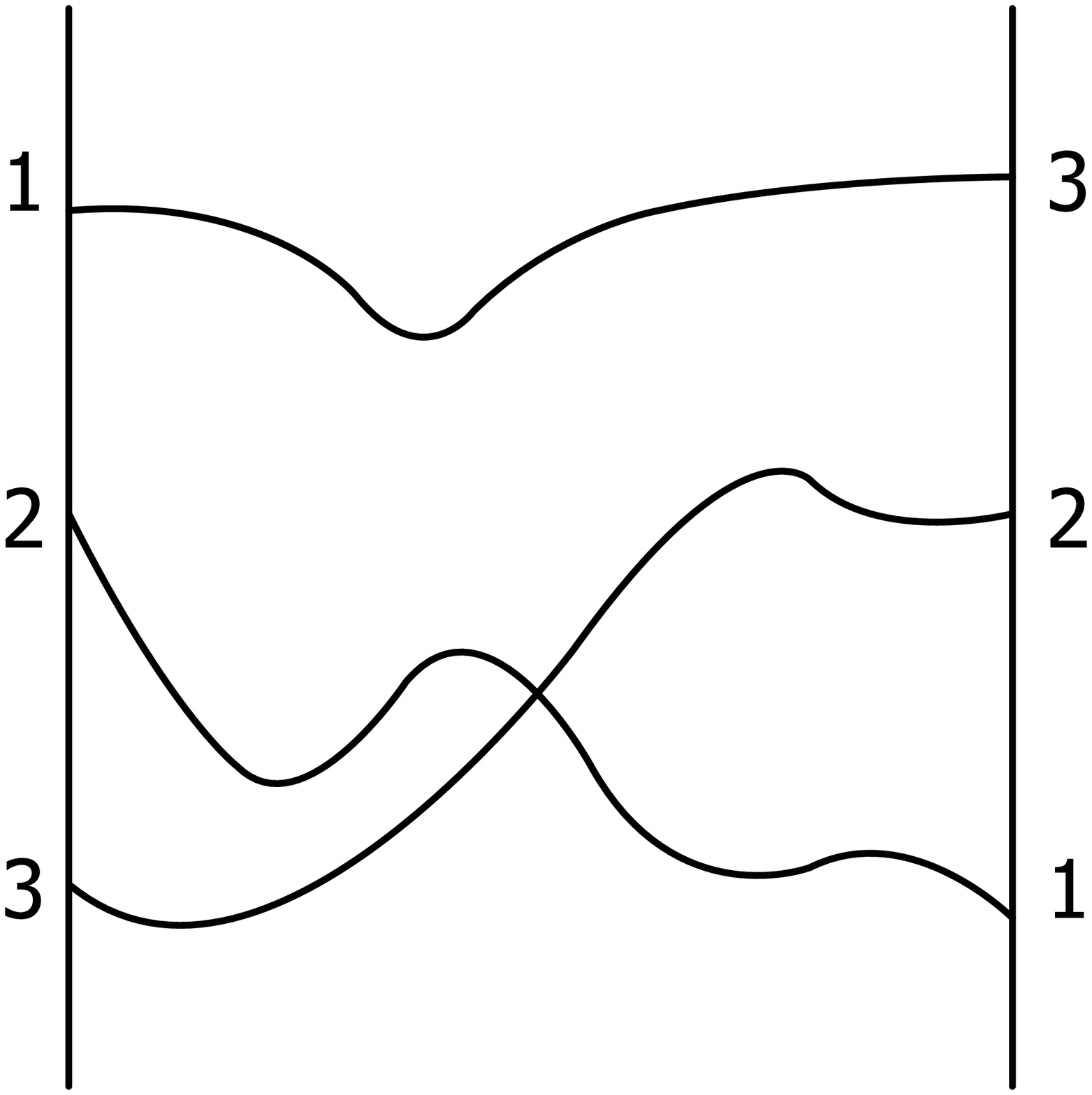}
\end{centering}
\caption{\label{3x3}All distinct  level diagrams for $N=3$
maximal Hamiltonians, $H(u)=T+uV$, drawn according to \re{rule}. There are $(N-1)!=2$ distinct diagrams each corresponding
to $N=3$ different orderings of eigenvalues of $V$ (see the text).
  For example, diagram a)
corresponds to $d_{1} >d_{2} >d_{3}$ and two other orderings obtained with a shift of the indices  by an integer
mod $N$, i.e. $d_{2} >d_{3} >d_{1}$ and
$d_{3} >d_{1} >d_{2}$. The diagrams
  predict   a single level crossing in agreement with \eref{crossnum} and specify which levels cross. Since this is
  also the maximum number of crossings for $N=3$, no multiple crossings of the same two levels are allowed, cf. Fig.~\ref{4x4}
  and \ref{fig: double cross}.}
\end{figure}

Now let us turn our attention to an arbitrary maximally commuting Hamiltonian $H(u)$. In the previous section we have
established the behavior of its energies  in $u\to\pm\infty$ limits. It follows from
\eref{hpr} that the energy level of $H(u)=T+uV$ (with appropriate rescaling \re{scale})
that starts from $-d_k$ at $u\to-\infty$ ends at $d_{k-1}$ at $u\to\infty$.
Symbolically, this can be represented by
\beg
k \longmapsto k-1,\quad (\mathrm{mod}\; N).
\label{trans}
\en
  Note that we cannot fix an ordering of
$d_k$ without loss of generality,   as the $d_k$ correspond to $\eps_k$,
see e.g. \eref{eH}, and we have already fixed the order of $\eps_k$ so that $\eps_1<\eps_2<\dots<\eps_N$. First,
we assume that all $d_k$ are distinct as is generally the case.
\eref{trans} implies that the flow of energy levels from $u=-\infty$ to $u=\infty$ can be schematically depicted
using the following rules:\\
\begin{enumerate}

\item Create two columns in which $\left\{ -d_{k}\right\} $ and $\left\{ d_{k}\right\} $
are both in descending order and replace each $d_{k}$ with its
lower index $k$, i.e.
\beg
\left.
\begin{array}{c}
-d_{i}\\
-d_{j}\\
\vdots\\
-d_{l}\\
-d_{m} \\
\end{array}\right|
\qquad\qquad
\left|\begin{array}{c}
d_{m}\\
d_{l}\\
\vdots\\
d_{j}\\
d_{i}
\end{array}\right.
 \Longrightarrow\left.
\begin{array}{c}
i\\
j\\
\vdots\\
l\\
m
\end{array}\right|
\qquad\qquad\left|
\begin{array}{c}
m\\
l\\
\vdots\\
j\\
i
\end{array}\right.
\label{rule}
\en

\item Draw a line connecting $j$ in the left column to $j-1$ in the right. These lines represent energy levels of $H(u)$.
 Consequently, their crossings imply crossings of the corresponding energy levels of $H(u)$.
\end{enumerate}
An example of an energy level diagram generated using the above prescription for $N=5$ is shown in Fig.~\ref{schem5}.
It corresponds to the ordering $d_{4} >d_{5} >d_{3} >d_{1} >d_{2}$ and predicts six level crossings. It also specifies
which levels cross, e.g. the top level connecting 2 and 1 crosses with the next in energy level connecting 1 and
5. We see that the crossing predicted by Fig.~\ref{schem5} are exactly the same as those of actual levels
of a maximally commuting
operator with that ordering shown in Fig.~\ref{real5}. The latter has been  obtained by  numerical diagonalization
of a $5\times 5$ maximally commuting operator \re{Hmel}, $H(u)$, with randomly chosen $\gamma_k$, $\varepsilon_k$
and random $d_k$ obeying the above ordering. More examples of level diagrams are shown in Figs.~\ref{maxcross}, \ref{4x4}, and
\ref{3x3} and discussed in detail below. Next, we explore further consequences of \eref{trans}.

\noindent {\bf Inevitability of level crossings for maximally commuting operators.} In Sec.~\ref{algebra} we have
seen that the maximum allowed number of level crossings is $(N-1)(N-2)/2$, see \eref{mmax}. Now let us show
that at least one crossing must be present in the spectrum of {\it any} maximally commuting $H(u)=T+uV$.
Suppose the eigenvalues of $V$ are ordered as $d_i<d_j<\dots<d_l<d_m$ as shown in the diagram \re{rule}
and assume there are no crossings.
Then, the top level must go from $i$ on the left   to $m$ on the right, i.e. $i\to m$,
the next level starting at $j$ on the left must be connected to $l$ on the right, $j\to l$ etc. Finally, we must have
$l\to j$ and $m\to i$. Consider in particular levels $i\to m$ and $m\to i$. According to \eref{trans},
this asymptotic behavior implies $m=i-1\mbox{ (mod $N$)}$ and $i=m-1\mbox{ (mod $N$)}$. We obtain $0=2\mbox{ (mod $N$)}$,
which does not hold for any $N\ge 3$, i.e. the above assumption that
levels do not cross cannot be true. Thus, we have demonstrated that at least one level crossing is always present.

The maximum number of crossings $M_c^{\max}=(N-1)(N-2)/2$ is realized e.g. for the ordering $d_1>d_2>\dots>d_N$, see Fig.~\ref{maxcross}.
In this case \eref{trans} implies $N\to N-1$ yielding $m=N-2$ crossings, $N-1\to N-2$ giving rise to another $m=N-3$ crossings
etc., so that altogether we have $\sum_{m=1}^{N-2}m=M_c^{\max}$ level crossings. Note however that the schematic level diagrams,
such as those shown in Figs.~\ref{schem5}, \ref{maxcross},  \ref{4x4}, and \ref{3x3}, do not account for the possibility of two levels crossing
more than once. For example, the level $2\to 1$ in Fig.~\ref{4x4}b  can go below the level $1\to 4$  and come back above it again generating two additional crossings, see Fig.~\ref{fig: double cross}. Therefore,
multiple crossings of two levels
can increase the total number of crossings $M_c$ for a given ordering of $d_k$ by an even number except when
$M_c=M_c^{\max}$. In the latter case, since $M_c$ cannot exceed $M_c^{\max}$, multiple crossings of the same two levels
are prohibited. We see that multiple crossings do not modify \eref{crossnum}.    Interestingly, numerically
we have found that for as low as $N=8$,  multiple crossings of
the same two levels are very common.

  Thus far, we have established that the total number of crossings $M_c$ in the spectrum of an arbitrary maximally commuting
Hamiltonian  is $1\le M_c \le M_c^{\max}$. By inspecting all level diagrams for $3\le N\le 6$, we have also found
that for a given $N$  the total number of level crossings changes in increments of 2 from $M_c^{\max}$
to 1 (2) for odd (even) $M_c^{\max}$, i.e. we verified \eref{crossnum} for these $N$. Moreover, this equation
is also supported by the parity considerations in the end of Sec.~\ref{algebra} and
is consistent with all preceding observations regarding the properties of level diagrams. As such, we
adopt it without  a formal proof.

Let us also comment that cases when some of the eigenvalues of $V$, $d_k$, are degenerate should be regarded
as crossings at $u\to\pm\infty$. Equivalently, one can treat $T$ and $V$ on equal footing by defining
$H(u,v)=vT+uV$. Then, degenerate $d_k$ correspond to crossings at $v/u=0$, while the crossings considered above
occur either at finite $v/u$ or at $v/u\to\pm\infty$, or equivalently at $u/v=0$. For example, levels of the BCS Hamiltonian, which is a linear combination of Gaudin magnets\cite{integr}, $\hat H_{BCS}=\frac{1}{B}\sum \eps_i \hat h^i(B)+\mbox{const}$,  cross at the value of the BCS coupling constant $g=1/B=1/u=\infty$ or, equivalently, at
$u=0$,  see e.g. Refs.~\onlinecite{Gaudin} and \onlinecite{Roman}.

We conclude this section with a discussion of useful properties and examples of energy level diagrams. There are
$N!$ diagrams for a given $N$ corresponding to permutations of eigenvalues $d_1,d_2,\dots,d_N$. However, some
of them are identical. Specifically, orderings $d_i<d_m<\dots<d_l$ and $d_{i+a}<d_{m+a}<\dots<d_{l+a}$ that differ
by a shift of indices by an integer $a$ yield identical diagrams, since \eref{trans}
is invariant with respect to the replacement $k\to k+a\mbox{ (mod $N$)}$. Because $N$
different orderings can be generated using this shift, it leaves $(N-1)!$ distinct diagrams.
For example, there are two distinct diagrams for $N=3$, see Fig.~\ref{3x3}. Each
 corresponds to three different orderings of $d_k$. Both diagrams predict a single level crossing.
Since this is also the maximum number of crossings for $N=3$, repeated crossings of the same two levels are not allowed.
Therefore, a single crossing of either two top or two bottom levels is the only option for $N=3$ maximally commuting
Hamiltonians. For $N=4$ there are six distinct level diagrams shown in Fig.~\ref{4x4}. Four of them -- diagrams a), c), d)
and f) in Fig.~\ref{4x4} -- exhibit the maximum number, $M_c^{\max}=3$, of level crossings. In a manner similar to that of the $N=3$ case
this is the only option for the corresponding sixteen orderings of $d_k$. In contrast, in diagrams b) and e) showing
a single crossing, multiple crossings can occur. This will increase the total number of crossings from one to three,
 see Fig.~\ref{fig: double cross}.

\section{Submaximal Hamiltonians}
\label{submax}

 The preceding
sections have focused on maximally commuting Hamiltonians, where we have explicitly constructed
these operators, solved them exactly, and used the solution to explain
the level crossings in such systems. In this section, we explore Hamiltonians linear in a parameter $u$
characterized by less than the maximum number of commuting partners. Most importantly, we demonstrate
that some of these submaximal Hamiltonians have {\it no} energy  level crossings, i.e. the {\it inevitability}
of level crossings due to parameter-dependent commuting partners appears to be an exclusive property of
maximal Hamiltonians.

As discussed in the Introduction, a given family of maximal
Hamiltonians contains $N-1$ nontrivial independent commuting
operators (see the discussion in the paragraph following \eref{arbH}). It is reasonable to expect that there exist submaximal
families with $N-2, N-3$ etc. Hamiltonians. Similar to \eref{nou},
any common $u$-independent symmetry is assumed to be factored out
by going to blocks of the same symmetry. We may adopt a convenient
terminology, where families with $N-L$ nontrivial Hamiltonians are
identified as being Type $L$ (cf.  Type I and II of
Ref.~\onlinecite{Shastry}). Then, the maximally commuting
Hamiltonians are Type 1, those with $N-2$ commuting operators are
Type 2 and so on. Since a nontrivial family must contain at least
two nontrivial commuting operators, the first nontrivial
instance of Type 1 occurs for $N=3$, Type 2 for $N=4$ etc., where
$N$ is the dimensionality of the state space.

\begin{figure}
\begin{centering}
\includegraphics[scale=0.85]{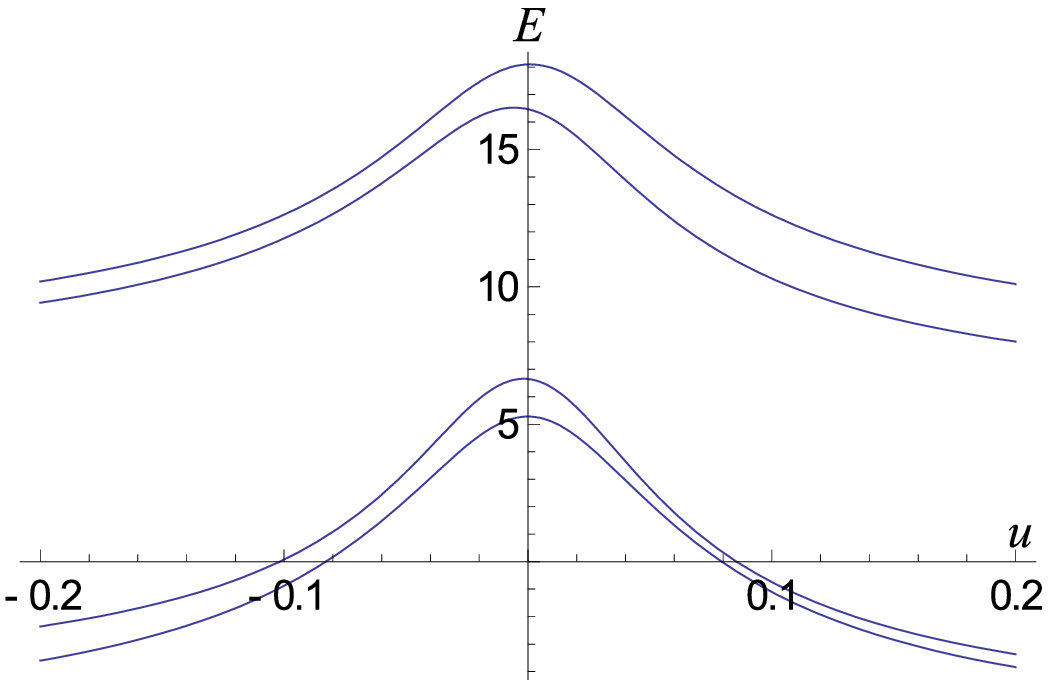}
\end{centering}
\caption{\label{4x4 no cross spec} Numerical energy levels of a  $4\times4$ {\it submaximal} Hamiltonian $H(u)$ obtained
from Eqs.~\re{eq: explicit param off diag double} and \re{type2}  with $x_0=1$,
random $y_0$, $d_{k}$, $\varepsilon_{k}$, and
$\protect\widetilde{\eps}_{k}$. Energies are scaled with  $(u^{2}+1)^{-1/2}$ as in Fig.~\ref{fig: double cross}.
Unlike $N=4$ maximal Hamiltonians, which always have two nontrivial commuting partners, this
$H(u)$ has only one such partner (see the text).
Note that levels of $H(u)$ do not cross at any $u$, i.e.  the mere existence of a nontrivial commuting partner
 does not guarantee  level crossings. This is to be contrasted with $4\times4$ maximally commuting Hamiltonians
 which always exhibit either three or one crossings, see Figs.~\ref{4x4} and \ref{fig: double cross}. }
\end{figure}

First, let us construct $4\times 4$ Type 2  Hamiltonians linear in a real parameter $u$.
Our task is therefore to identify two $4\times 4$ commuting real symmetric matrices that do not have
the third independent commuting partner other than $(a+ub)I$. We will do so by employing the construction
of maximal Hamiltonians detailed in Sec.~\ref{sec:Type-I-Basis}.
Consider  $4\times4$ operators $H\left(x,y,u\right)=x\, T+y\, K+u\, V$
and $\widetilde{H}\left(x,y,u\right)=x\, \widetilde{T}+y\, \widetilde{K}+u\, \widetilde{V}$, linear
in parameters $x$, $y$ and $u$, such that
\begin{equation}
\left[H\left(x,y,u\right),\widetilde{H}\left(x,y,u\right)\right]=0.
\label{eq: bilinear commute}
\end{equation}
Since this equation is to hold for all $x$, $y$, and $u$, the coefficients of the $xy$, $xu$, $yu$  etc. terms must vanish individually.
We obtain
 \begin{equation}
\left[T,\widetilde{T}\right]=\left[V,\widetilde{V}\right]=\left[K,\widetilde{K}\right]=0,
\label{triple}
\end{equation}
 \begin{equation}
\left[T,\widetilde{V}\right]=\left[\widetilde{T},V\right],\qquad\left[T,\widetilde{K}\right]=\left[\widetilde{T},K\right],
\label{trivial}
\end{equation}
 \begin{equation}
\left[V,\widetilde{K}\right]=\left[\widetilde{V},K\right].
\label{special}
\end{equation}
Let us choose these real symmetric matrices so that $(T+uV, \widetilde{T}+u\widetilde{V})$ and
$(K+uV, \widetilde{K}+u\widetilde{V})$ belong to two different families of maximally commuting Hamiltonians
parameterized by $\gamma_m, \eps_m$ and $\widetilde{\gamma}_m=\gamma_m, \widetilde{\eps}_m$, respectively, see
\eref{Hmel}. Then,   Eqs.~\re{triple} and \re{trivial} are satisfied by construction.  Consequently, it  remains to solve
\eref{special}.

Thus, from \eref{Hmel} we have in the common eigenbasis of $V$ and $\widetilde{V}$
\begin{equation}
\begin{array}{ccc}
\dis\left[H\left(x,y,u\right)\right]_{mn} & = & \dis x\,\gamma_{m}\gamma_{n}\dfrac{d_{m}-
d_{n}}{\varepsilon_{m}-\varepsilon_{n}}+y\,\gamma_{m}\gamma_{n}\dfrac{d_{m}-d_{n}}{\widetilde{\eps}_m-
\widetilde{\eps}_n},\qquad m\neq n,\\
\\
\dis \left[H\left(x,y,u\right)\right]_{mm} & = & \dis u\, d_{m} -x\sum_{j\neq m}\gamma_{j}^{2}
\dfrac{d_{m}-d_{j}}{\varepsilon_{m}-\varepsilon_{j}}-y\,\sum_{j\neq m}
\widetilde{\eps}_m\gamma_{j}^{2}\dfrac{d_{m}-d_{j}}{\widetilde{\eps}_m -\widetilde{\eps}_j},\\
\end{array}
\label{eq: explicit param off diag double}
\end{equation}
 where $d_{k}$  are the eigenvalues of  $V$. Matrix elements of $\widetilde{H}\left(x,y,u\right)$ are obtained
from \eref{eq: explicit param off diag double} by replacing $d_{k}\to \widetilde{d}_k$.
Using these expressions for the matrix elements, one can rewrite the remaining commutation relation \re{special}
as follows
\begin{equation}
\gamma_{l}^{2}=\frac{\left|\begin{array}{ccc}
1 & 1 & 1\\
d_{i} & d_{j} & d_{k}\\
\widetilde{d}_{i} & \widetilde{d}_{j} & \widetilde{d}_{k}
\end{array}\right|
\left|\begin{array}{ccc}
1 & 1 & 1\\
\varepsilon_{i} & \varepsilon_{j} & \varepsilon_{k}\\
\widetilde{\varepsilon}_{i} & \widetilde{\varepsilon}_{j} & \widetilde{\varepsilon}_{k}
\end{array}\right|^{2}}
{\left(\varepsilon_{i}-
\varepsilon_{j}\right)\left(\varepsilon_{j}-\varepsilon_{k}\right)\left(\varepsilon_{i}-\varepsilon_{k}\right)
\left(\widetilde{\varepsilon}_{i}-\widetilde{\varepsilon}_{j}\right)\left(\widetilde{\varepsilon}_{j}-\widetilde{\varepsilon}_{k}\right)
\left(\widetilde{\varepsilon}_{i}-\widetilde{\varepsilon}_{k}\right)},\quad l\neq i,j,k.
\label{eq: complicated}
\end{equation}

 Therefore, choosing $d_r, \widetilde{d}_r, \eps_r$, and $\widetilde{\varepsilon}_{r}$, we obtain $\gamma_r$ from
 \eref{eq: complicated}. This
 yields two commuting matrices $H(x,y,u)$ and $\widetilde{H}(x,y,u)$.
Fixing nonzero values of $x=x_0$ and $y=y_0$, we obtain a Type 2 family of Hamiltonians linear in $u$,
\beg
H\left(u\right)=\left(x_0\, T+y_0\, K\right)+u\, V,\quad
\widetilde{H}\left(u\right)=(x_0\, \widetilde{T}+y_0\, \widetilde{K})+u\, \widetilde{V}.
\label{type2}
\en
There are a number of equivalent ways to verify that these operators are indeed Type 2 rather than maximally commuting.
For example, one can show that their matrix elements \re{eq: explicit param off diag double} cannot be cast into the
form \re{Hmel}. Alternatively, it can be demonstrated that conditions \re{munu} necessary for
any maximal operator do not hold. However, a less formal, but more fruitful verification uses  the following
argument.
We have seen in Sec.~\ref{sec: Crossings} that {\it any} $N=4$ maximal Hamiltonian must have either one or three level crossings.
Let us check if this holds for the Hamiltonians \re{type2}. To this end, we set $x_0=1$, generate random
$y_0$, $d_k$, $\varepsilon_{k}$, and $\protect\widetilde{\eps}_{k}$, and obtain $\gamma_k$ from \eref{eq: complicated}
and $H(u)$ from \eref{eq: explicit param off diag double}. Doing so repeatedly and numerically diagonalizing the resulting
Hamiltonians we observe that they always have either no or two level crossings. An example with no crossings
is shown in Fig.~\ref{4x4 no cross spec}. Thus, operators \re{type2} are Type 2.

We  see that level crossings are not guaranteed when the number
of commuting operators is less than the maximum --  nontrivial solutions of \eref{comrel}
do not necessarily imply crossings. The converse is also  false, i.e. level crossings
can occur in the absence of any nontrivial commuting partner linear in $u$ and any $u$-independent symmetry.
 For example, one can show that the  $4\times4$ Hamiltonian given by \eref{xnocom} in the Appendix
with a single level crossing at $u=0$ has no nontrivial commuting partners and no
  $u$-independent symmetry. Interestingly, $N=4$ is the
first dimensionality where this happens as for $3\times3$ real symmetric  matrices linear in
 $u$ a level crossing implies a nontrivial commuting
partner  linear in $u$ and vice versa \cite{emil}.

\section{Summary and open questions}

In this paper, we addressed the problem of the  violation of the Wigner-von Neumann
non-crossing rule in quantum integrable systems. For this purpose, we introduced and
studied a general class of maximal Hamiltonians -- a vector space   of $N\times N$ real symmetric Hamiltonians, $H(u)=T+uV$,
characterized by the existence of the maximum possible number ($N$) of independent
mutually commuting integrals similarly linear in the coupling $u$. We have resolved this commutation property and
explicitly constructed general maximal Hamiltonians, see \eref{Hmel}. Interestingly, these operators are
equivalent to the Gaudin magnets \re{gaud1}  in the  next
to highest weight sector, $J^z=J^z_{\max}-1$, where $J^z$ is the
$z$ projection of the total spin.

The mapping to Gaudin magnets allowed us to obtain a
complete exact solution for the eigenstates and eigenvalues of $H(u)$, Eqs.~\re{xm} and \re{eH}. Furthermore,
we have demonstrated that energy level crossings are inevitable for maximal operators, i.e. there is always
at least one crossing. The total number of crossings varies from 1 or 2 to $(N-1)(N-2)/2$, see Sec.~\ref{sec: Crossings}.
Thus, the mere existence of the maximum number of commuting partners guarantees  a) an exact solution and
b) level crossings. This relationship between the existence of conservation laws and exact solution
 is a quantum analog of the famous Liouville-Arnold theorem in classical mechanics. The latter
states that if a classical model with $n$ degrees of freedom has $n$ Poisson-commuting integrals, its
equations of motion are exactly solvable\cite{arnold}.

At the same time, by constructing an explicit example
we have demonstrated that submaximal Hamiltonians -- real symmetric operators of the form $T+uV$ with
less than the maximum number of linear in $u$ commuting partners -- often show no instances of level crossings at any $u$.
Thus, the inevitability of crossings is an exclusive feature of  maximally commuting operators. Similarly, we have
also shown that the presence of level crossings does not necessarily imply the existence of a nontrivial commuting
partner linear in $u$, i.e. crossings can occur even in the absence of such partners as well as  $u$-independent symmetries.

Our understanding of properties of parameter-dependent energy spectra in integrable models is far from complete.
We conclude this section with a list of open questions stemming from the results of this work.

\begin{enumerate}

\item   We have shown that there are submaximal Hamiltonians with no level crossings. Nevertheless,
 crossings  often do occur in these systems in violation of the non-crossing rule, see  Sec.~\ref{submax}.
 This  indicates that there is more to the link between crossings and the presence of commuting
 partners. It is interesting to understand this link for submaximal operators, what lifts level
 repulsion in this case, and why crossings happen only for a fraction of submaximal Hamiltonians.

\item  In Sec.~\ref{submax}, we have also introduced a notion of Type $L$ commuting family characterized
by $N-L$ nontrivial integrals. In this classification
maximally commuting operators are Type 1, while submaximal operators correspond to $L\ge 2$. A natural
question is whether there is a general explicit parametrization  for Type 2, 3 etc. similar to the one obtained
in this paper for maximal systems. For instance, one can show that $|J^z|\le J^z_{\max}-2$ sectors
provide examples of submaximal commuting families.

\item What is the role of maximal Hamiltonians in the context of general quantum  integrable  Hamiltonians?
For the central spin Hamiltonians (Gaudin magnets) they represent the next to highest weight sector.   Do other integrable models have maximally commuting sectors?

\item  In this paper, we focused   on operators linear in the coupling $u$. An interesting
question is how our results can  be generalized to operators with a more general, e.g. polynomial, dependence on the
coupling.

\item We have established hard bounds for the number $M_c$ of level crossings
in $N\times N$ maximally commuting operators.  Can one also determine the distribution of $M_c$
for large $N$, i.e. the relative prevalence of maximal  Hamiltonians with a
particular number of level crossings?

\end{enumerate}

\section{Acknowledgements}

This research was partially supported by the National Science Foundation award NSF-DMR-0547769 (H. K. Owusu and
E. A. Yuzbashyan).
E. A. Yuzbashyan also acknowledges the  financial support by a David and Lucille Packard Foundation Fellowship for Science and Engineering, and Alfred P. Sloan Research Fellowship.

\appendix

\section{Shastry's construction of commuting matrices}
\label{Shastry's construction}

In Ref.~\onlinecite{Shastry} Shastry developed an approach to generate commuting real symmetric $N\times N$ matrices
linear in a parameter $u$. Here we show that matrices obtained with this approach belong to maximally commuting
set constructed in Sec.~\ref{sec:Type-I-Basis}, see \eref{Hmel}.

First, we briefly review the results of Ref.~\onlinecite{Shastry}. Consider \eref{comrel}. In the common
eigenbasis of $V$ and $\widetilde{V}$ the second relation in \eref{comrel} becomes
\begin{equation}
S_{ij}\equiv\dfrac{T_{ij}}{d_{i}-d_{j}}=\dfrac{\widetilde{T}_{ij}}{\widetilde{d}_{i}-\widetilde{d}_{j}},\quad i\ne j,
\label{ss}
\end{equation}
where $T_{ij}$  ($\widetilde{T}_{ij}$) are the matrix elements of $T$ ($\widetilde{T}$) and $d_i$ and
$\widetilde{d}_{i}$ are the eigenvalues of $V$ and $\widetilde{V}$, respectively. It remains to consider
the $[T,\tilde{T}]=0$ commutation relation in \eref{comrel}. This can be cast into the following form:
\begin{equation}
\mu_{ijk}\tilde{d}_{i}+\mu_{jki}\tilde{d}_{j}+\mu_{kij}\tilde{d}_{k}+
\sum_{l\neq i,j,k}\nu_{lijk}\tilde{d}_{l}=0,\quad \mbox{distinct $i,j,k$,}
\label{eq: Constraint}
\end{equation}
 where $\mu_{ijk}$ and $\nu_{lijk}$ depend only on matrix elements of $H(u)$ and not on those of $\widetilde{H}(u)$.
 Specifically, they involve only  $d_{r},T_{rr}$ and $S_{rm}$.

A  set of particular solutions to \eref{comrel} can be obtained by setting the coefficients in \eref{eq: Constraint}
at each $\widetilde{d}_r$ individually to zero, i.e.
\begin{equation}
\mu_{ijk} =  0,\quad\nu_{lijk}=0,\quad \mbox{distinct $l,i,j,k$,}
\label{munu}
\end{equation}
Now commuting $H(u)$ and $\widetilde{H}(u)$ can be generated as follows. One first chooses $3N-1$ parameters, e.g.
$2N-3$ variables $S_{1r}$ and $S_{2r}$ for $\{2,3\}\le r\le N$ and $N+2$ variables  $\{d_r\}$, $T_{11}$, and $T_{22}$.
Then, \eref{munu} reduce to linear equations and can be solved for the remaining variables. Once $H(u)$ is determined in this
way, $\widetilde{H}(u)$ can also be constructed, see Ref.~\onlinecite{Shastry} for details. This scheme is quite suitable
for numerical implementation and, having examined several examples, Shastry
observed crossings in all cases. Based on this and the results of Ref.~\onlinecite{emil}
he conjectured that these matrices will always exhibit them.

To show that this construction always yields maximal Hamiltonians, we note that \eref{munu} is a sufficient
condition for \eref{eq: Constraint} to have $N$ linearly independent solutions for
$(\widetilde{d}_1, \widetilde{d}_2,\dots\widetilde{d}_N)$. Since $\widetilde{d}_r$ are the eigenvalues
of $\widetilde{V}$, the existence of $N$ linearly independent solutions means that there are $N$ Hamiltonians $\widetilde{H}(u)$
with linearly independent $\widetilde{V}$s. The absence of $u$-independent symmetries can also be demonstrated (it follows
from $S_{ij}\ne 0$ for all $i\ne j$). Thus, we have a maximally commuting set, see the Introduction. The only difference
is that by construction $d_r$ are not allowed to be degenerate, see \eref{ss}, while the maximal set contains these matrices
as well.

Finally, we  write down an example (see the discussion at the end of Sec.~\ref{submax})
 of a $4\times 4$ Hamiltonian $H(u)$ with a level crossing but no $u$-independent symmetry and
no commuting partners linear in $u$ other than trivial ones -- $cH(u)+(a+ub)I$, where $a$, $b$, and $c$ are real numbers and $I$ is
the identity matrix,
\beg
H(u)= \left(\begin{array}{cccc}
1 & 0 & 0 & 0\\
0 & -1 & 0 & 0\\
0 & 0 & 0 & 0\\
0 & 0 & 0 & 0\end{array}\right)+u\left(\begin{array}{cccc}
1 & -1 & 1 & -1\\
-1 & 1 & -1 & 1\\
1 & -1 & -2 & 1\\
-1 & 1 & 1 & 2\end{array}\right).
\label{xnocom}
\en


\begin{thebibliography}{99}
\bibitem{aside} We note that the restriction to Hermitian operators in the definition of the maximal
set is not essential for
this construction. Specifically, following the above arguments, one can show that an arbitrary maximally commuting  and not necessarily Hermitian operator can be written in the form $\Sigma H(u) \Sigma^{-1}$, where $\Sigma$ is
diagonal and $H(u)$ is a symmetric matrix parametrized as in \eref{Hmel} but  with arbitrary complex parameters $\Sigma_{jj}, d_j, \gamma_{j}$, and $\varepsilon_{j}$.
\bibitem{note2} Any $u$-independent symmetry common to all $H^{i}\left(u\right)$
can always be removed by going to smaller blocks.
\bibitem{note1} This can be seen by evaluating $mn^{\mathrm{th}}$ matrix elements of  $\protect\sum_{k=1}^N f_{k}(u)\,h^{k}(u)=0$
for $m\ne n$ and using Eqs.~\re{hmel} and \re{eq: sum to identity}.
\end{thebibliography}
\end{document}